\documentclass[lettersize,journal]{IEEEtran}
\usepackage{amsmath,amsfonts,amssymb}
\usepackage{algorithmic}
\usepackage{algorithm}
\usepackage{array}
\usepackage[caption=false,font=normalsize,labelfont=sf,textfont=sf]{subfig}
\usepackage{textcomp}
\usepackage{stfloats}
\usepackage{url}
\usepackage{verbatim}
\usepackage{graphicx}
\usepackage{cite}
\usepackage{todonotes}
\usepackage{svg}
\usepackage{siunitx}
\usepackage{multirow}
\usepackage{acro} %Abbreviations
\DeclareAcronym{BEV}{
  short = BEV,
  long = battery electric vehicle,
  long-plural-form = battery electric vehicles,
}

\DeclareAcronym{ADAS}{
  short = ADAS,
  short-plural-form = ADAS,
  long = advanced driver assistance system,
  long-plural-form = Advanced Driver Assistance Systems,
}

\DeclareAcronym{ACC}{
  short = ACC,
  long = Adaptive Cruise Control,
  long-plural-form = N/A
}

\DeclareAcronym{UCC}{
  short = UCC,
  long = Urban Cruise Control,
  long-plural-form = N/A
}

\DeclareAcronym{UEQ}{
  short = UEQ,
  long = User Experience Questionnaire,
  long-plural-form = N/A
}

\DeclareAcronym{HMI}{
  short = HMI,
  long = Human Machine Interface,
  long-plural-form = N/A
}

\DeclareAcronym{LSA}{
  short = LKA,
  long = Lane Keeping Assistant,
  long-plural-form = N/A
}

\DeclareAcronym{CID}{
  short = CID,
  long = Central Information Display,
  long-plural-form = N/A
}

\DeclareAcronym{VR}{
  short = VR,
  long = Virtual Reality,
  long-plural-form = N/A
}

\DeclareAcronym{MISC}{
  short = MISC,
  long = Misery Scale,
  long-plural-form = N/A
}

\DeclareAcronym{NDRT}{
  short = NDRT,
  short-plural-form = NDRT,
  long = Non-Driving Related Task,
  long-plural-form = Non-Driving Related Tasks
}

\DeclareAcronym{VDL}{
  short = VDL,
  long = VanDerLaan,
  long-plural-form = N/A
}

\DeclareAcronym{DALI}{
  short = DALI,
  long = Driver-Activity-Load-Index,
  long-plural-form = N/A
}

\DeclareAcronym{DMS}{
  short = DMS,
  short-plural-form = DMS,
  long = Driver Monitoring System,
  long-plural-form = Driver Monitoring Systems
}

\DeclareAcronym{TOR}{
  short = TOR,
  long = Takeover Request,
  long-plural-form = Takeover Requests
}
\DeclareAcronym{ICC}{
  short = ICC,
  long = Intraclass Correlation Coefficient,
  long-plural-form = N/A
}
\DeclareAcronym{FDR}{
  short = FDR,
  long = False Discovery Rate,
}
\DeclareAcronym{SHAP}{
  short = SHAP,
  long = SHapley Additive exPlanations,
}
\DeclareAcronym{LOSO}{
  short = LOSO,
  long = Leave-One-Subject-Out,
}
\usepackage{pifont}
\newcommand{\cmark}{\ding{51}} % check mark
\newcommand{\xmark}{\ding{55}} % cross mark

\newif\ifanonymize
\anonymizefalse % Original
\newcommand{\anonymtext}[2]{%
  \ifanonymize
    \textit{#2}%
  \else
    #1%
  \fi
}

\begin{document}

\title{Driver Behavior Under Traffic Complexity: Variance Decomposition and Cross-Driver Generalization for Driver Monitoring}

\author{Lukas Köning, Nataša Miličić and Klaus Bogenberger%
  %\thanks{Manuscript received [TO BE FILLED]; revised [TO BE FILLED]; accepted [TO BE FILLED]. \textit{(Corresponding author: Lukas Köning)}}%
  \thanks{This work involved human subjects or animals in its research. Approval of all ethical and experimental procedures and protocols was granted by BMW's ethics committee, and performed in line with the Declaration of Helsinki.}%
  \thanks{Lukas Köning is with the Chair of Traffic Engineering and Control, Technical University of Munich, 80333 Munich, Germany and the BMW AG, 80809 Munich, Germany \textit{(e-mail: lukas.koening@tum.de)}.}%
  \thanks{Nataša Miličić is with the BMW AG, 80809 Munich, Germany.}%
  \thanks{Klaus Bogenberger is with the Chair of Traffic Engineering and Control, Technical University of Munich, 80333 Munich, Germany.}%
}

% The paper headers
%\markboth{IEEE Transactions on Intelligent Vehicles,~Vol.~[TO BE FILLED], No.~[TO BE FILLED], July~2026}%
\markboth{}%
{Köning \MakeLowercase{\textit{et al.}}: Driver Behavior Under Traffic Complexity: Variance Decomposition and Cross-Driver Generalization for Driver Monitoring}

\IEEEpubid{This work has been submitted to the IEEE for possible publication. Copyright may be transferred without notice, after which this version may no longer be accessible.}
% \IEEEpubid{0000--0000/00\$00.00~\copyright~2021 IEEE}
% Remember, if you use this you must call \IEEEpubidadjcol in the second
% column for its text to clear the IEEEpubid mark.

\maketitle

\begin{abstract}
Understanding why a traffic situation is demanding for the human driver is central to safe and comfortable partially automated driving. Existing complexity metrics characterize only external environmental factors, while driver monitoring systems detect only endogenous states such as distraction. Neither side captures how external demands translate into driver-perceived situational load. Driver behavior serves as the natural bridge between both, and this paper investigates the inverse inference problem of estimating traffic complexity from behavioral signals. A systematic screening of 175 behavioral features across five domains (gaze, head pose, longitudinal control, guiding fixation, scanning strategy) is applied to data from 20~drivers in real urban traffic. Of these, 31~features exhibit statistically confirmed complexity effects, while 140~are confirmed as null effects through equivalence testing. Mixed-effects variance decomposition reveals that complexity explains only 1.5\% of behavioral variance, whereas driver identity accounts for 23\% and residual variance for 75\%. This unfavorable ratio explains both the failure of all eight evaluated feature-level personalization strategies and the convergence of four classification architectures at F1 around 0.45 under leave-one-subject-out cross-validation. Guiding fixation rate emerges as the single most deployment-ready feature, combining speed-robustness, universality across drivers, and minimal inter-driver variation in complexity sensitivity. The results define three deployment regimes for complexity-adaptive advanced driver assistance systems and establish the variance structure as the primary bottleneck for complexity estimation.
\end{abstract}


\section{Introduction}
% Intro + Problem statement
\IEEEPARstart{T}{he} cognitive demands placed on a driver using partially automated \ac{ADAS} (SAE Level 2) in urban traffic range from passive monitoring on empty arterials to active multi-agent coordination at complex intersections. Yet this variation in driver demand with traffic complexity remains invisible to modern \acp{DMS} and \ac{ADAS}, even as these systems become increasingly common in production vehicles~\cite{Chalmers2025}. Partially automated \ac{ADAS} assume lateral and longitudinal control, but the driver must still supervise the system~\cite{SAEJ3016}. Accordingly, UN Regulation No.~171 on partially automated \ac{ADAS} mandates \acp{DMS} for such systems, and vehicle rating institutions award credits for advanced \ac{DMS} functionality, such as phone usage and drowsiness detection~\cite{DCAS2024,EuroNCAP2026}.

Maintaining driver engagement is challenging because many systems rely on warnings triggered by detected inattention, which drivers often perceive as punitive~\cite{Bazilinskyy2015}. Current \ac{DMS} detect the endogenous driver state but not the exogenous situational demands imposed by traffic complexity. This gap limits cooperative, context-aware \ac{ADAS}, as alert drivers in complex situations receive the same system response as those on an empty road.
While drivers generally adopt more defensive driving behaviors as complexity rises, current state-of-the-art \ac{ADAS} do not adapt to driver-perceived situational demands~\cite{Xie2021_2,Lyu2017}. 
Therefore, understanding the current demand placed on the driver and incorporating it into a complexity-adaptive \ac{ADAS} is crucial for providing optimal \ac{ADAS} that promote active participation and avoid overload. 
% Possible adaptations include varying driving dynamics during maneuvers, reducing \ac{HMI} input in complex scenarios, and improving \ac{TOR} timing. 
% Increasing drivers' engagement in the driving task by developing \ac{ADAS} that promote active participation is a promising approach to enhancing road safety~\cite{Illgner2024,Koening2026}.
\IEEEpubidadjcol{}
% \todo[inline]{IEEEpubidadjcol aktivieren}

%Whether an external element imposes situational load depends not only on its existence but also if the driver perceives it as demanding. Behavior encodes this human assessment and therefore carries information that purely environmental metrics cannot provide.
Whether external factors impose situational load depends on whether the driver perceives them as demanding, and behavior encodes this human assessment, carrying information that purely environmental metrics cannot provide.
Estimating traffic complexity from environmental features is a well-established research field~\cite{Liu2024,Zhang2021}. However, the inverse inference problem of deriving complexity from driver behavior, and thus from the driver's own situational assessment, is sparsely studied, even though the required sensor setup (\ac{DMS}) is already mandated in production vehicles. While a few studies have examined behavioral differences across complexity levels, including our own preceding work identifying 16 metrics that differentiate among three expert-labeled complexity levels, these analyses share several limitations~\cite{Wang2023,Halin2025,Kunst2025,Koening2026ITSC}. Feature spaces are narrow and selected ad hoc rather than systematically screened, speed confounds are uncontrolled, variance decomposition into complexity-driven versus driver-idiosyncratic components is absent, and neither the cross-driver generalization problem nor the effectiveness of per-driver normalization at the feature level has been addressed.
% Furthermore, neither the cross-driver generalization problem, i.e., whether complexity can be inferred from a previously unseen driver, nor the effectiveness of per-driver normalization at the feature level has been addressed.
%Furthermore, the cross-driver generalization problem, whether complexity can be inferred from a previously unseen driver, as required for deployment, has not been addressed. Lastly, the question of whether per-driver normalization can mitigate inter-driver differences at the feature level has not been evaluated. This paper addresses these gaps by answering the following research questions:

% Feature spaces are narrow and selected ad hoc rather than systematically screened, speed confounds are not controlled, and the decomposition of observed effects into complexity-driven versus driver-idiosyncratic variance is not attempted. Furthermore, the cross-driver generalization problem, whether complexity can be inferred from a previously unseen driver, as required for deployment, has not been addressed. Lastly, the question of whether per-driver normalization can mitigate inter-driver differences at the feature level has not been evaluated. This paper addresses these gaps by answering the following research questions:

\begin{itemize}
    \item RQ1: Which driver-behavior features are sensitive to traffic complexity, and which are not?
    \item RQ2: How much of the observed feature variance is explained by traffic complexity versus stable driver identity?
    \item RQ3: Can per-driver normalization recover complexity signal that is masked by inter-driver differences?
    \item RQ4: To what extent can complexity be inferred from driver behavior when the driver is previously unseen?
\end{itemize}

% This paper advances the state of the art by offering novel, systematic insights into driver behavior under traffic complexity in real-world urban traffic with partially automated \ac{ADAS}.
This paper contributes a systematic screening of 175 behavioral features across five domains (Kruskal-Wallis, TOST, speed residualization), a mixed-effects variance decomposition separating complexity from driver identity, a factorial evaluation of eight feature-level personalization strategies, and a cross-driver classification benchmark across four architectures under \ac{LOSO} cross-validation and block-aware cross-validation.
% The main contributions of this paper are summarized as follows:
% \begin{itemize}
%     \item A systematic screening of 175 driver behavior metrics across five behavioral domains, distinguishing genuine complexity effects from null effects through rigorous statistical testing (Kruskal-Wallis, TOST, speed control).
%     \item Quantification of the relative contributions of traffic complexity, stable driver identity, driver-specific complexity sensitivity, and residual variance using mixed-effects variance decomposition (random intercept and random slope).
%     \item Testing whether per-driver personalization at the feature level can recover the complexity signal through a systematic evaluation of eight feature-level personalization strategies.
%     \item Quantification of the cross-driver generalization gap across four model architectures under leave-one-subject-out and block-aware cross-validation.
% \end{itemize}
Together, these contributions provide an end-to-end analysis framework, from feature screening through variance explanation to classification, that links the statistical structure of driver-behavior features to design constraints for complexity-adaptive \ac{DMS} and \ac{ADAS}.

% Following this introduction, Section \ref{sec:Related_Work} summarizes related work; Section \ref{sec:Methods} introduces the methodology; Section \ref{sec:Results} presents the results, which are discussed in Section \ref{sec:Discussion}. Conclusions and implications for future work are presented in Section \ref{sec:Conclusion}.
\section{Related Work}
\label{sec:Related_Work}
\subsection{Traffic Complexity}
Despite widespread use in recent traffic safety, human factors, and automated driving research, the concept of traffic complexity lacks a consensus definition \cite{Zhou2025_2}. Publications use different terms, e.g., traffic scene complexity, traffic situation complexity, and environmental complexity \cite{Liu2024,ScharfeScherf2021_2,Yang2021}. For simplicity, this paper uses \textit{traffic complexity} to encompass all concepts, knowingly disregarding minor differences. Conceptually, traffic complexity is understood as an exogenous property of the traffic situation, independent of the driver's individual characteristics. Specifically, it denotes the demand imposed on the driver by factors such as road geometry, traffic interactions, and environmental conditions. Traffic complexity is related to, but distinct from, mental workload: complexity describes the external task demand, whereas workload refers to the internal cognitive cost of meeting that demand~\cite{Fuller2000,Paxion2014}. % A high-complexity situation increases task demand, which may or may not translate into elevated workload depending on the driver's capability and coping strategies. 
This paper targets complexity as the exogenous construct. Existing approaches to quantify traffic complexity can be separated into objective approaches that derive traffic complexity from measurable features of the traffic situation and subjective approaches that rely on cognitive models and expert judgment. The following paragraphs review both.

\subsubsection{Objective Quantification}
Objective approaches derive complexity from measurable features of the traffic situation, typically distinguishing static elements (road geometry, traffic signs, weather) from dynamic elements (vehicles, pedestrians, cyclists) \cite{Yang2021,Cheng2022,Liu2024}. Methods range from weighted sums and entropy-based measures to potential field models and deep learning \cite{Liu2024,Cheng2022,Zhang2021}. The primary use cases are scenario selection for \ac{ADAS} testing and online complexity estimation for behavior adaptation \cite{Liu2024,Zhang2021}. While criticality is sometimes used as a proxy, a complex situation need not be critical \cite{Westhofen2023}. A fundamental limitation shared across objective approaches is that ground-truth labels are either expert-rated or derived from downstream algorithm performance, leaving no independently annotated complexity benchmark. Recent work attempts to bridge objective and subjective evaluations, but a standardized benchmark remains absent \cite{Cao2025}.

\subsubsection{Subjective Quantification}
The closest theoretical anchor to subjective traffic complexity is Fuller's Task-Capability Interface (TCI) model, which characterizes driving as a dynamic control task in which the outcome balances task demand, i.e., the objective complexity imposed by road, traffic, and environment, and driver capability \cite{Fuller2000}. The complexity of the current traffic situation strongly influences task demand, particularly for perception tasks \cite{Paxion2014}. Teh et al. \cite{Teh2014} demonstrated that complexity fluctuates over time during a drive and affects both subjective workload ratings and objective driving performance. Drivers tend to drive more defensively, and increased complexity reduces lane-keeping performance and raises the probability of errors~\cite{Xie2021_2,Lyu2017,Paxion2014}. In the context of automated driving, research has found declining take-over performance as complexity increases \cite{ScharfeScherf2021_2}.
Although the effects of traffic complexity on the human driver are well known and actively researched, the subjective quantification of traffic complexity remains restricted to questionnaires and expert labeling \cite{Teh2014,Boelhouwer2020}. Fastenmeier and Gstalter~\cite{Fastenmeier2007} operationalized the TCI concept through driving task analysis, decomposing routes into subtasks and assigning ordinal complexity ratings by expert judgment, an approach that forms the basis for expert-labeling protocols.
The behavioral response to traffic complexity has received comparatively little attention in the literature. Beyond the previously mentioned longitudinal control adaptations, only a few publications investigate other driver behavior metrics, such as gaze and head tracking, even though they are used as state of the art in driver state detection by \ac{DMS}. Prior research has identified wider, more dispersed gaze behavior in more complex scenarios \cite{Kunst2025,Halin2025,Wang2023}. Our previously published analysis, which this paper extends, revealed significant patterns across 16 behavior metrics covering longitudinal control, gaze metrics, fixation frequency, and guiding fixation behavior \cite{Koening2026ITSC}.

Overall, subjective complexity quantification remains limited to questionnaires and expert ratings, and the inverse inference problem of estimating complexity from driver behavior remains sparsely studied.
% Overall, while objective quantification has been extensively studied, subjective quantification remains limited to questionnaires and expert ratings. Subjective traffic complexity is rarely linked to feature-based behavioral classification. The inverse inference problem of estimating complexity from driver behavior rather than from infrastructure or expert ratings remains sparsely studied.

\subsection{Driver Behavior Features for Task Demand and Complexity Estimation}
Ohn-Bar and Trivedi \cite{OhnBar2016} survey visual cues for intelligent vehicles and distinguish endogenous driver state monitoring (fatigue, distraction) from exogenous context understanding (traffic complexity), noting that the latter is mostly addressed via infrastructure sensors rather than driver behavior. Current \ac{DMS} research focuses on endogenous states using gaze and head pose features, achieving high accuracy under controlled conditions but degrading with driver heterogeneity \cite{AlQuraishi2024}.

For exogenous task demand, drivers produce more exploratory scanning patterns as demand increases \cite{Doshi2012}. Simulator studies reported increased vertical gaze dispersion in complex situations \cite{Kunst2025,Halin2025}, while real-world lane-change studies found more dispersed glances \cite{Wang2023}. Information-theoretic gaze metrics complement distributional measures: Shiferaw et al. \cite{Shiferaw2019} distinguish stationary gaze entropy (spatial dispersion) from gaze transition entropy (sequential predictability), with the former being more robust to inter-driver heterogeneity \cite{Goodridge2024}. Sample entropy captures temporal irregularity and increases with cognitive load \cite{Bakhchina2025}, though Arutyunova et al. \cite{Arutyunova2025} question whether entropy reflects experience rather than task demands. Head pose provides complementary but coarser attention information \cite{Jha2023,Mikula2020,OhnBar2016}, and research has identified driver-idiosyncratic patterns that mask complexity variance \cite{Mikula2020,Jha2023}. Scanning strategy features (fixation frequency, scanpath length, head-gaze coupling) capture how drivers distribute visual attention, with exploration increasing under complexity \cite{Shiferaw2019,Doshi2012,OhnBar2016}. Physiological measures track task demand with higher accuracy but require intrusive sensors incompatible with production vehicles \cite{AlQuraishi2024}.

Guiding fixation couples gaze behavior with the driving task, defined as fixating the tangent point and trajectory about 2~seconds ahead \cite{Land1994,Lappi2022}. These patterns persist in assisted driving \cite{Mole2021}, but apart from our preceding publication, no prior work examines complexity effects on guiding fixation \cite{Koening2026ITSC}. 
Longitudinal control features reflect complexity-adapted behavior: drivers increase braking and reduce speed as complexity rises \cite{Xie2021_2,Lyu2017}, and behavioral entropy of control inputs has been validated for distraction detection \cite{Boer2001,Chang2025}.

No prior study applies such a comprehensive multi-domain screening with variance decomposition to separate complexity from driver identity, and most prior work targets generic task demands rather than traffic complexity specifically.
% No prior study applies a comprehensive screening across gaze, head pose, longitudinal control, guiding fixation, and scanning strategy within the same dataset, with variance decomposition to separate complexity from driver identity. Furthermore, most prior work targets generic task demands rather than traffic complexity specifically.

\subsection{Cross-Driver Generalization and ADAS Personalization}
When driver monitoring models are evaluated in \ac{LOSO} settings, performance drops substantially compared with within-driver evaluation for both distraction and drowsiness detection \cite{GLi2024}. Pe\~{n}a et al. \cite{Pena2025} report that physiological clustering before model training partially mitigates this gap, but their approach requires intrusive sensors. The underlying cause of stable inter-driver differences in baseline behavior has been quantified via \ac{ICC} analysis for isolated driving decisions, e.g., overtaking maneuvers \cite{Rasch2024}. However, no study has systematically applied \ac{ICC} decomposition across a multi-domain feature set to explain the cross-driver gap.

To address inter-driver heterogeneity, personalization strategies operate at multiple levels. At the feature level, per-driver z-score normalization removes between-subject offsets and is common in affective computing \cite{Mariooryad2015}. Per-driver baseline subtraction has been used in driver behavior analysis \cite{Krejtz2018,Dong2025}. At the model level, domain adaptation transfers a pretrained model to a target driver without labeled data \cite{GLi2024}, while few-shot and meta-learning approaches require only a small calibration set \cite{Lu2023}. These strategies are typically evaluated on binary state detection tasks (e.g., attentive versus distracted) and have not been tested on the more subtle signal of traffic complexity, where effect sizes are considerably smaller.

No prior work formally decomposes the variance sources that cause the cross-driver gap into complexity-driven, driver-identity, and residual components, nor evaluates whether feature-level personalization can recover a complexity signal masked by inter-driver differences.
\section{Methods}
\label{sec:Methods}
\begin{figure}[!htbp]%
    \centering
    %\includesvg[width=0.45\textwidth, height=\textheight, keepaspectratio]{figures/Pipeline.svg}
    \includegraphics[width=0.45\textwidth, height=\textheight, keepaspectratio]{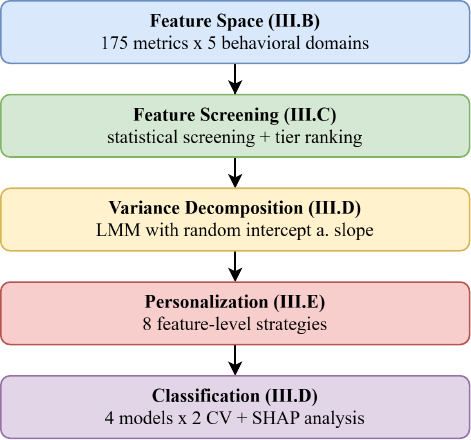}
    \caption{Overview of the analysis pipeline of this paper.}
    \label{fig:pipeline}
\end{figure}

The analysis follows a four-stage pipeline, illustrated in Figure~\ref{fig:pipeline}: (1)~systematic feature screening across five behavioral domains with speed-confound control, (2)~variance decomposition via mixed-effects \ac{ICC} analysis, (3)~feature-level personalization testing, and (4)~cross-driver classification benchmarking with post-hoc \ac{SHAP} analysis~\cite{Lundberg2017}.

\subsection{Dataset and Experimental Setup}
\label{sec:Dataset}
The analysis is based on data from a real-vehicle study conducted in urban traffic in \anonymtext{Munich, Germany}{an anonymous Central European city}, described in detail in \cite{Koening2026,Koening2026ITSC}. From the original 57~participants, a subsample of 20~drivers ($\mu = 38.8$~years, $\sigma = 10.2$~years, 3~female, 17~male) was selected via stratified sampling. Continuous expert labeling of synchronized front-camera video requires approximately 3× real-time effort per drive. Therefore, labeling all 57 drives was not feasible within the  available annotation budget. Stratified sampling across age and driving experience ensured maximum demographic coverage within this constraint. The resulting 20-group design provides adequate power for fixed-effects estimation but limits precision of random-slope covariance estimates (Section~\ref{sec:Disc_Limitations}). All participants declared informed consent.

For the real-vehicle study, \anonymtext{a pre-series BMW i5 M60}{an anonymous car} was used. The vehicle was equipped with \anonymtext{the 2021 version of BMW's}{anonymized} production-grade \ac{ACC} and \ac{LSA}, classifying it as a partially automated \ac{ADAS}~\cite{SAEJ3016}. A specialized software version enabled fully shared longitudinal control, allowing the driver to brake without deactivating the system~\cite{Illgner2024,Koening2026}. For the analysis in this paper, only the first drives of participants were selected, during which the \ac{ADAS} was characterized by fully-shared longitudinal control. Therefore, only segments with active \ac{ADAS} were analyzed, as fully-shared longitudinal control resulted in nearly 100\% \ac{ADAS} usage. Participants were informed in advance that they would use a partially automated \ac{ADAS} that assists with maintaining a safe distance from preceding vehicles and staying in lane. Participants were instructed to use the \ac{ADAS} as often as possible, provided they felt comfortable doing so. 
The route, displayed in Figure~\ref{fig:Route}, spanned approximately 11~km of multi-lane, single-lane, and side roads in urban traffic, covering diverse scenarios including construction sites, intersections, and varying traffic densities. Drives were conducted between 9~am and 5~pm during summer, resulting in varying traffic, lighting, and weather conditions.

\ifanonymize
\begin{figure}[!htbp]
    \centering
    \textit{Figure removed due to anonymization. Image contains route with relevant traffic situations marked as pictograms (e.g., traffic signs).}
    \caption{Route driven in the real-vehicle study.}
    \label{fig:Route}
\end{figure}
\else
\begin{figure}[!htbp]
    \centering
    \includegraphics[width=0.35\textwidth, height=\textheight, keepaspectratio]{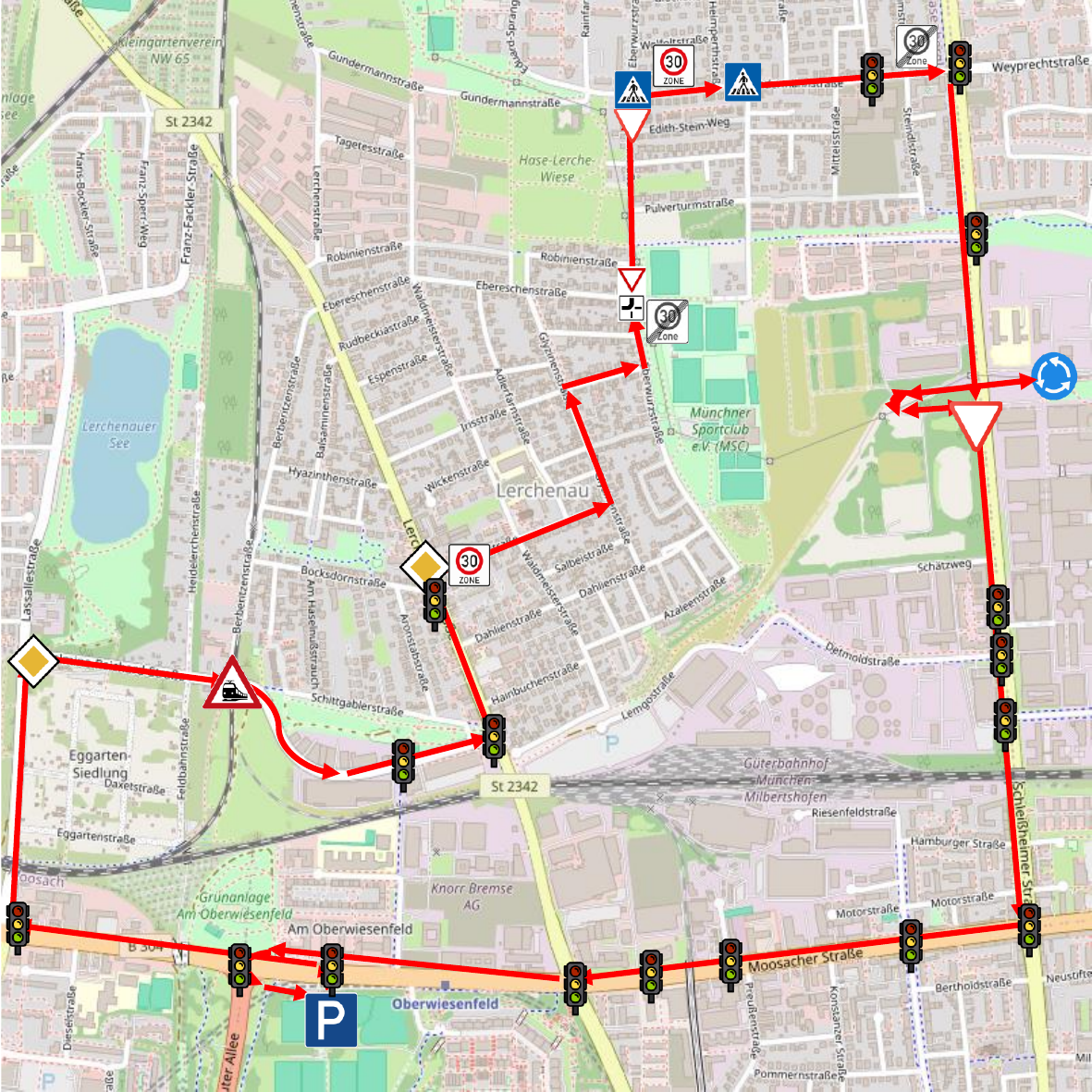}
    \caption{Route driven in the real-vehicle study.}
    \label{fig:Route}
\end{figure}
\fi

Driver behavior data were extracted from the vehicle's control units at a native sampling rate between 20 and 100~ms and linearly interpolated onto a fixed 100~ms grid. The extracted dataset includes GPS position, vehicle speed, \ac{ADAS} activation state, brake pedal usage, and \ac{DMS} signals (gaze yaw, gaze pitch, head yaw, head pitch, head roll) from the vehicle's camera-based \ac{DMS} located behind the steering wheel. The total dataset comprises approximately 400,000 samples (11~hours cumulative driving) after filtering periods where vehicle speed was below 1~km/h or the driver was not looking outside the vehicle windows.

Three experts covering automotive engineering, traffic psychology, and usage safety independently labeled each drive using three ordinal complexity classes: \textit{simple} (clearly structured, low attention demand), \textit{normal} (partially structured, minor adjustments needed), and \textit{complex} (multifaceted, frequent adjustments, heightened attention). Experts labeled complexity based solely on the traffic scene (road layout, traffic participants, infrastructure) visible in synchronized front-camera video, without access to driver behavior data. Labeling was performed time-continuously without predefined segments using a dedicated software tool~\cite{Koening2026ITSC}. A $\pm$2~s tolerance window was applied to account for reaction time variability in label-change timestamps~\cite{Mariooryad2015}. A sensitivity analysis was conducted to assess the impact of temporal tolerance ($\pm$0~s to $\pm$3~s) on inter-rater reliability. Krippendorff's $\alpha$ increased from 0.50 at $\pm$0~s to 0.57 at $\pm$2.0~s (+14.3\%). Beyond $\pm$2.0 s, improvements followed a pattern of diminishing returns, at the cost of temporal precision required for event-level analysis. Temporal stability, measured by Spearman's $\rho$, remained high across all tolerances ($\rho \geq 0.83$), confirming that the aggregated labels are robust to the choice of window size. Based on this trade-off between reliability and temporal precision, $\pm$2~s was selected for all subsequent analyses, matching typical human reaction/annotation latency ($\sim$1-2~s)~\cite{Mariooryad2015}.

Individual labels were aggregated using the Dawid-Skene algorithm~\cite{DawidSkene1979}. The resulting class distribution is 27\% simple, 45\% normal, and 28\% complex. Inter-rater reliability yields Krippendorff's $\alpha = 0.57$ (moderate agreement), while individual rater agreements with the combined labels reach substantial values ($\kappa = 0.68$--$0.82$), justifying the labels as expert-aggregated proxy labels of subjective traffic complexity~\cite{LandisKoch1977}. More details on the labeling can be found in the preceding conference publication \cite{Koening2026ITSC}.

\subsection{Feature Space Definition}
\label{sec:FeatureSpace}
The 175 candidate metrics arise from a systematic cross-product of 35 base signals, organized into five behavioral domains, and a standard derivation set. This design ensures uniform coverage: every base signal is characterized by the same statistical descriptors, producing directly comparable features across domains.

Each base signal $s(t)$ is expanded into up to six time-series derivations computed over causal, right-aligned rolling windows (3~s at 10~Hz): (D1)~raw value $s(t)$, (D2)~rolling mean $M(s(t))$, (D3)~rolling standard deviation $SD(s(t))$ (spherical standard deviation for angular signals), (D4)~rolling Shannon entropy $H_{Sh}(s(t))$, (D5)~rolling sample entropy $H_{Sa}(s(t))$ ($m = 2$, $r = 0.2\,SD$), and (D6)~first temporal derivative (gradient).
Base signals that are themselves kinematic derivatives (velocity, acceleration) receive a reduced set (D2-D5). Combined 2D dispersion signals receive D3-D5 only. Pre-smoothed signals, i.e., those that are themselves rolling means or continuous rates (e.g., braking intensity as a 15~s rolling mean of the binary brake signal), receive the full set D1--D6 except for D2, as a rolling mean of an already-smoothed signal trivially correlates with D1.

The 35 base signals span five behavioral domains. \textit{Gaze Behavior} (10~signals, 45~metrics) covers camera-based gaze yaw~$\Psi_g$ and pitch~$\theta_g$ at 10~Hz, their velocities and accelerations, combined 2D dispersion~$g_{comb}$, and blink rate~$f_{blink}$. \textit{Head Pose} (9~signals, 42~metrics) comprises 3D head orientation (yaw, pitch, roll) with velocity and acceleration derivatives. \textit{Longitudinal Control} (4~signals, 24~metrics) includes braking acceleration during active braking~$a_{brake}$, time- and distance-based braking intensity~$BR_{t/s}$, and speed-limit deviation~$\Delta v$. \textit{Guiding Fixation} (6~signals, 36~metrics) captures the guiding fixation rate~$GFR$ (proportion of time the driver's gaze falls within $\pm$10\textdegree{} of the trajectory tangent point at a speed-dependent look-ahead distance), forward and backward polling rates, angular deviation from the guiding fixation point~$\delta_{GF}$, and time since last guiding fixation~$t_{GF}$. \textit{Scanning Strategy} (6~signals, 36~metrics) covers fixation frequency~$f_{fix}$, scanpath length~$SPL$, look-ahead distance, head--gaze coupling~$\rho_{hg,\theta}$ (yaw and pitch), and extreme head pose rate. % The complete specification of all base signals and their derivations is provided in supplementary Table~S3 (\ref{tab:feature_space}).

\subsection{Feature Screening}
\label{sec:Screening}
The screening pipeline operates in two stages: (1)~sample-level statistical screening with \ac{FDR} correction, speed-residual control, and TOST equivalence testing, and (2)~driver-level tier ranking to address temporal autocorrelation.

For each of the 175 metrics, the non-parametric Kruskal-Wallis $H$ test is applied to test whether complexity levels differ. Features are retained when effect size $\eta^2 \geq 0.01$ (Cohen's~\cite{Cohen1988} small effect-size benchmark) and $p_{\text{adj}} < 0.05$ (Benjamini-Hochberg~\cite{BenjaminiHochberg1995} \ac{FDR} correction across all 175 tests). The $\eta^2 \geq 0.01$ threshold reflects that complexity produces subtle behavioral modulations, with $N \approx 400$k, post-hoc power exceeds 0.99 for this effect size, so features failing the threshold genuinely lack practical relevance. To isolate complexity-driven effects from speed-mediated effects, OLS residualization is applied to each retained metric $y$:
\begin{equation}
    y_{\text{res}} = y - (\hat{\beta}_0 + \hat{\beta}_1 \cdot \text{speed})
    \label{eq:speed_residual}
\end{equation}
where $\hat{\beta}_0$ and $\hat{\beta}_1$ are estimated via ordinary least squares. Kruskal-Wallis is then re-run on $y_{\text{res}}$. Features with $\eta^2_{\text{res}} \geq 0.01$ are classified as \textit{speed-robust}, those with $\eta^2_{\text{res}} < 0.01$ as \textit{speed-mediated}. For all metrics that fail the initial Kruskal-Wallis test ($\eta^2 < 0.01$), Two-One-Sided Tests (TOST) with equivalence bounds of $\Delta = \pm 0.2$ pooled standard deviations (Cohen's $d = 0.2$) are applied to formally confirm absence of a practically significant effect~\cite{Lakens2017}.

The sample-level Kruskal-Wallis test operates on $N \approx 400$k samples at 10~Hz, violating the independence assumption. To verify that effects survive aggregation, each feature is aggregated to per-driver per-complexity-level medians ($N=60$) and the Kruskal-Wallis test is re-run. Features with $\eta^2_{\text{driver}} \geq 0.01$ and $p_{\text{driver}} < 0.05$ are classified as Tier~1 (robust) and those with $\eta^2_{\text{driver}} \geq 0.01$ but $p_{\text{driver}} \geq 0.05$ as Tier~2 (power-limited, retained because the effect is present but statistical power is insufficient at $N =60$). Features with $\eta^2_{\text{driver}} < 0.01$ are dropped (Tier~3). Ranking stability is assessed via Spearman correlation between $\eta^2_{\text{sample}}$ and $\eta^2_{\text{driver}}$.

\subsection{Variance Decomposition}
\label{sec:VarDecomp}
For each retained feature (Tier~1 + Tier~2), a linear mixed-effects model with random intercept and random slope is fitted:
\begin{equation}
    y_{ij} = \beta_0 + \beta_1 \cdot x_{ij} + u_{0j} + u_{1j} \cdot x_{ij} + \varepsilon_{ij}
    \label{eq:mixed_model}
\end{equation}
where $i$ indexes samples, $j$ indexes drivers, $x \in \{0, 1, 2\}$ codes complexity ordinally, $u_{0j}$ is the random intercept (driver baseline), $u_{1j}$ is the random slope (driver-specific complexity sensitivity), and $\varepsilon_{ij} \sim \mathcal{N}(0, \sigma^2_\varepsilon)$~\cite{Agresti2010}.

Variance is decomposed following the Nakagawa and Schielzeth~\cite{Nakagawa2013} framework into three components: marginal $R^2_m$ (variance explained by complexity), \ac{ICC} ($\sigma^2_{u0} / \sigma^2_{\text{total}}$, proportion attributable to stable driver identity), and slope share ($\sigma^2_{u1} / (\sigma^2_{u0} + \sigma^2_{u1})$, proportion of between-driver variance reflecting differences in complexity \textit{sensitivity} versus baseline level). Features with $\text{ICC} < 0.5$ are classified as \textit{universal} (consistent across drivers), those with $\text{ICC} \geq 0.5$ as \textit{idiosyncratic} (dominated by driver identity). Because $\sigma^2_{\text{total}}$ includes the intercept-slope covariance term, a strongly negative covariance (drivers with higher baselines showing weaker complexity sensitivity) can reduce the denominator below $\sigma^2_{u0}$, yielding ICC values exceeding unity for features with pronounced negative intercept-slope coupling. A dummy-coding sensitivity check confirmed coding-invariance (median $\Delta R^2_m = 0.001$, Spearman $\rho = 0.70$, $p < 0.001$).

\subsection{Personalization Strategies}
\label{sec:Personalization}
To test whether feature-level normalization can recover complexity signal from inter-driver variance, eight normalization strategies plus a raw baseline~(L0) are evaluated in a $2 \times 4$ factorial design. The two normalization types are z-score (center~+~scale; L1--L4) and mean-subtraction (center only; M1--M4). Each type is crossed with four temporal scopes: full-drive (L1/M1, non-causal), calibration phase using only the first 3~minutes (L2/M2, causal), expanding window from drive start with 60~s warmup (L3/M3, causal), and sliding window over the preceding 5~minutes (L4/M4, causal and driver-agnostic). Each strategy is applied independently per feature before recomputing~$\eta^2$. Effectiveness is measured as $\Delta \eta^2$ relative to L0.

\subsection{Classification Pipeline}
\label{sec:Classification}
Four model architectures are evaluated: Gradient Boosting (GB, HistGradientBoosting with per-fold tuning of depth, learning rate, and number of estimators), Support Vector Machine (SVM, per-fold grid search over $C$ and $\gamma$), a 2-layer GRU with temporal attention (128 units, 3~s input windows, dropout 0.1, early stopping), and Logistic Regression. Additionally, two reference baselines are included: an unsupervised PCA baseline (PC1 projected onto the test set, sign-flipped for positive label correlation, and discretized into three classes via equal-frequency quantile binning) and majority-class prediction. All features use causal rolling windows, i.e., no future information enters the computation.

Two cross-validation strategies disentangle inter-driver and within-driver generalization. \ac{LOSO} trains on 19 drivers and tests on the held-out driver's complete drive, serving as the \textit{primary evaluation} because it mimics deployment on a previously unseen driver. BlockGroupKFold (5-fold, stratified by driver, within-driver block shuffling) evaluates within-driver generalization while respecting temporal structure. Hyperparameter tuning is performed via per-fold grid search on the training set. Macro F1 is reported to account for 3-class imbalance.

\ac{SHAP} values are computed using TreeExplainer on the GB model under \ac{LOSO} (up to 2,000 test samples per fold, concatenated across all 20 folds). The global importance metric is mean~$|\text{SHAP}|$ per feature, averaged over samples and classes. Unlike univariate screening ($\eta^2$), \ac{SHAP} reflects multivariate predictive importance and captures redundancy, complementarity, and interaction effects.

\section{Results}
\label{sec:Results}

\subsection{Feature Screening}
\label{sec:Results_Screening}
The two-stage screening pipeline reduces 175 candidate features to 31 retained metrics. In Stage~1, 32 features pass the Kruskal-Wallis test ($\eta^2 \geq 0.01$, $p_{\text{adj}} < 0.05$). Table~\ref{tab:screening_variance} lists all 32 features grouped by behavioral domain, together with the variance decomposition results discussed in Section~\ref{sec:Results_VarDecomp}. ICC values exceeding unity reflect strongly negative intercept-slope covariance (see Section~\ref{sec:VarDecomp}). Detailed test statistics are provided in supplementary Table~\ref{tab:screening_stats}.

% TABLE 1 (combined screening + variance decomposition)
% ============================================================
% TABLE 1: Combined Feature Screening & Variance Decomposition
% ============================================================
\begin{table*}[!t]
\centering
\caption{Feature screening and variance decomposition results of all metrics passing Kruskal-Wallis screening. Rob.: speed-robust~(\cmark) or speed-mediated~(\xmark). Tier: 1~=~robust at driver level, 2~=~power-limited, 3~=~not driver robust.}
\label{tab:screening_variance}
%\scriptsize
%\setlength{\tabcolsep}{2pt}
\begin{tabular}{c ccc cc c c ccc}
\multirow{2}{4em}{\textbf{Feature}} & \textbf{Simple} & \textbf{Normal} & \textbf{Complex} & \multirow{2}{1em}{\textbf{$\eta^2$}} & \multirow{2}{2em}{\textbf{$\eta^2_{res}$}} & \multirow{2}{1.5em}{\textbf{Rob.}} & \multirow{2}{1.5em}{\textbf{Tier}} & \multirow{2}{2em}{\textbf{$R^2_m$}} & \multirow{2}{2em}{\textbf{ICC}} & \multirow{2}{2em}{\textbf{Slope}} \\
 & \textbf{$\mu~(\sigma)$} & \textbf{$\mu~(\sigma)$} & \textbf{$\mu~(\sigma)$} & \\
\hline
  \textbf{Longitudinal Control} \\
  $BR_t$\,[\si{\percent}] & 2.79 (10.64) & 4.86 (10.96) & 11.02 (17.49) & 0.057 & 0.036 & \cmark & 1 & 0.002 & 1.161 & 0.211 \\
  $H_{Sa}(BR_t)$\,[$-$] & 0.01 (0.05) & 0.03 (0.07) & 0.06 (0.10) & 0.046 & 0.031 & \cmark & 1 & 0.047 & 0.011 & 0.604 \\
  $H_{Sh}(BR_t)$\,[$-$] & 0.21 (0.68) & 0.32 (0.76) & 0.60 (0.97) & 0.033 & 0.021 & \cmark & 2 & 0.033 & 0.020 & 0.396 \\
  $M(a_{brake})$\,[\si{\meter/\second\squared}] & -1.23 (0.86) & -1.05 (0.66) & -0.88 (0.62) & 0.030 & 0.036 & \cmark & 2 & 0.001 & 0.251 & 0.242 \\
  $a_{brake}$\,[\si{\meter/\second\squared}] & -1.33 (0.92) & -1.23 (0.78) & -0.99 (0.73) & 0.028 & 0.019 & \cmark & 2 & 0.001 & 0.102 & 0.812 \\
  $M(\Delta v)$\,[\si{\percent}] & -21.65 (30.78) & -30.94 (30.72) & -34.98 (30.16) & 0.028 & 0.006 & \xmark & 1 & 0.025 & 0.044 & 0.463 \\
  $\Delta v$\,[\si{\percent}] & -21.82 (30.59) & -31.33 (31.52) & -34.81 (31.09) & 0.026 & 0.009 & \xmark & 1 & 0.023 & 0.043 & 0.450 \\
  $BR_s$\,[\si{\percent}] & 3.83 (6.16) & 5.17 (6.47) & 6.93 (9.20) & 0.025 & 0.013 & \cmark & 1 & 0.025 & 0.309 & 0.470 \\
  $H_{Sa}(\Delta v)$\,[$-$] & 0.71 (0.69) & 0.59 (0.64) & 0.47 (0.50) & 0.021 & 0.006 & \xmark & 1 & 0.021 & 0.017 & 0.478 \\
  $SD(BR_t)$\,[\si{\percent}] & 0.56 (2.26) & 0.75 (1.97) & 1.41 (2.79) & 0.020 & 0.013 & \cmark & 1 & 0.019 & 0.018 & 0.385 \\
  $H_{Sa}(BR_s)$\,[$-$] & 0.48 (0.72) & 0.61 (0.76) & 0.74 (0.79) & 0.016 & 0.015 & \cmark & 1 & 0.014 & 0.064 & 0.196 \\
  $H_{Sh}(\Delta v)$\,[$-$] & 0.95 (0.97) & 1.09 (0.95) & 1.27 (0.93) & 0.015 & 0.004 & \xmark & 1 & 0.015 & 0.013 & 0.502 \\
  $SD(\Delta v)$\,[\si{\percent}] & 2.89 (3.85) & 3.32 (3.79) & 3.94 (3.92) & 0.010 & 0.003 & \xmark & 1 & 0.008 & 0.005 & 0.708 \\
  \hline
  \textbf{Gaze Behavior} \\
  $f_{blink}$\,[\si{1/\minute}] & 17.49 (31.93) & 24.28 (42.40) & 32.96 (48.55) & 0.019 & 0.017 & \cmark & 2 & 0.022 & 0.262 & 0.113 \\
  $H_{Sh}(\Psi_g)$\,[$-$] & 1.54 (0.80) & 1.76 (0.82) & 1.77 (0.83) & 0.016 & 0.004 & \xmark & 1 & 0.015 & 0.099 & 0.107 \\
  $H_{Sh}(\dot{\Psi}_g)$\,[$-$] & 1.13 (0.68) & 1.31 (0.67) & 1.33 (0.68) & 0.016 & 0.005 & \xmark & 2 & 0.011 & 0.178 & 0.051 \\
  $SD(g_{comb})$\,[\si{\radian}] & 0.15 (0.03) & 0.15 (0.02) & 0.15 (0.02) & 0.015 & 0.014 & \cmark & 2 & 0.018 & 0.960 & 0.147 \\
  $SD(\Psi_g)$\,[\si{\radian}] & 0.14 (0.03) & 0.15 (0.03) & 0.15 (0.03) & 0.015 & 0.014 & \cmark & 2 & 0.011 & 1.340 & 0.156 \\
  $H_{Sh}(\ddot{\Psi}_g)$\,[$-$] & 1.08 (0.61) & 1.23 (0.61) & 1.25 (0.63) & 0.014 & 0.004 & \xmark & 1 & 0.010 & 0.140 & 0.061 \\
  $H_{Sh}(g_{comb})$\,[$-$] & 2.51 (0.80) & 2.70 (0.82) & 2.69 (0.84) & 0.011 & 0.003 & \xmark & 2 & 0.017 & 0.239 & 0.233 \\
  \hline
  \textbf{Head Pose} \\
  $SD(\Psi_h)$\,[\si{\radian}] & 0.11 (0.02) & 0.12 (0.02) & 0.12 (0.02) & 0.020 & 0.018 & \cmark & 2 & 0.019 & 1.376 & 0.254 \\
  $H_{Sh}(\ddot{\Psi}_h)$\,[$-$] & 1.31 (0.54) & 1.37 (0.52) & 1.46 (0.57) & 0.010 & 0.006 & \xmark & 2 & 0.015 & 0.811 & 0.192 \\
  \hline
  \textbf{Guiding Fixation} \\
  $GFR$\,[\si{\percent}] & 77.01 (22.76) & 73.17 (22.58) & 66.31 (24.94) & 0.029 & 0.022 & \cmark & 1 & 0.017 & 0.468 & 0.033 \\
  $H_{Sh}(\delta_{GF})$\,[$-$] & 1.14 (0.75) & 1.33 (0.77) & 1.41 (0.78) & 0.018 & 0.004 & \xmark & 1 & 0.016 & 0.076 & 0.115 \\
  $H_{Sh}(\delta_{GF,out})$\,[$-$] & 0.75 (0.80) & 0.93 (0.87) & 1.05 (0.90) & 0.017 & 0.005 & \xmark & 1 & 0.017 & 0.094 & 0.105 \\
  $H_{Sa}(\delta_{GF,out})$\,[$-$] & 0.17 (0.25) & 0.21 (0.27) & 0.25 (0.30) & 0.011 & 0.004 & \xmark & 1 & 0.006 & 0.310 & 0.028 \\
  $H_{Sh}(t_{GF})$\,[$-$] & 0.09 (0.28) & 0.13 (0.34) & 0.18 (0.39) & 0.011 & 0.005 & \xmark & 3 & --- & --- & --- \\
  \hline
  \textbf{Scanning Strategy} \\
  $SPL$\,[\si{\degree}] & 306.65 (145.43) & 348.90 (147.78) & 344.57 (152.48) & 0.016 & 0.024 & \cmark & 2 & 0.013 & 0.425 & 0.074 \\
  $M(SPL)$\,[\si{\degree}] & 303.04 (148.39) & 343.31 (150.12) & 339.33 (153.71) & 0.014 & 0.025 & \cmark & 2 & 0.011 & 0.405 & 0.072 \\
  $M(\rho_{hg,\theta})$\,[$-$] & 0.08 (0.32) & 0.09 (0.31) & 0.16 (0.33) & 0.012 & 0.011 & \cmark & 2 & 0.000 & 0.522 & 0.299 \\
  $\rho_{hg,\theta}$\,[$-$] & 0.07 (0.33) & 0.09 (0.32) & 0.16 (0.34) & 0.011 & 0.010 & \cmark & 2 & 0.000 & 0.416 & 0.302 \\
  $f_{fix}$\,[\si{\hertz}] & 2.49 (1.49) & 2.82 (1.37) & 2.71 (1.36) & 0.010 & 0.006 & \xmark & 2 & 0.005 & 0.216 & 0.042 \\
  \hline
  \textbf{\textit{Median (Tier 1+2)}} & & & & \textit{0.016} & \textit{0.011} & & & \textit{0.015} & \textit{0.228} & \textit{0.196} \\
\end{tabular}
\end{table*}

Longitudinal control dominates with the highest hit rate (13 of 24 candidates) and the four strongest effects ($\eta^2 = 0.028$--$0.057$). Guiding fixation rate ($GFR$, $\eta^2 = 0.029$) is the strongest non-longitudinal-control feature. Entropy-based derivations (D4, D5) account for 14 of 32 relevant features, while the gradient derivation (D6) is universally not relevant, formally confirmed via TOST. % --- all 22 gradient features across all domains fail screening, formally confirmed via TOST.
Of the 32 features, 17 are speed-robust ($\eta^2_{\text{res}} \geq 0.01$): all eight braking features retain their effect after speed control, whereas the five speed-deviation metrics do not. Among gaze features, all spherical standard deviation metrics are speed-robust. For the 143 features failing screening ($\eta^2 < 0.01$), TOST equivalence testing confirms 140 as genuine null effects (supplementary Table~\ref{tab:tost_null}). Key structural findings in the null set include: (a)~all pitch-axis signals (gaze and head, 28 derivations total) are null, while yaw-axis signals carry complexity information; (b)~forward and backward polling rates are null, while the continuous guiding fixation rate is among the strongest features.

In Stage~2, features are aggregated to per-driver per-complexity-level medians. Of the 32 features, 16 are Tier~1 (robust), 15 Tier~2 (power-limited), and one dropped (Tier~3: $H_{Sh}(t_{GF})$). Among speed-robust features, sample-level and driver-level rankings show strong agreement (Spearman $\rho = 0.618$, $p = .008$).

\subsection{Variance Decomposition}
\label{sec:Results_VarDecomp}
\begin{figure}[!htbp]
    \centering
    \includegraphics[width=\columnwidth]{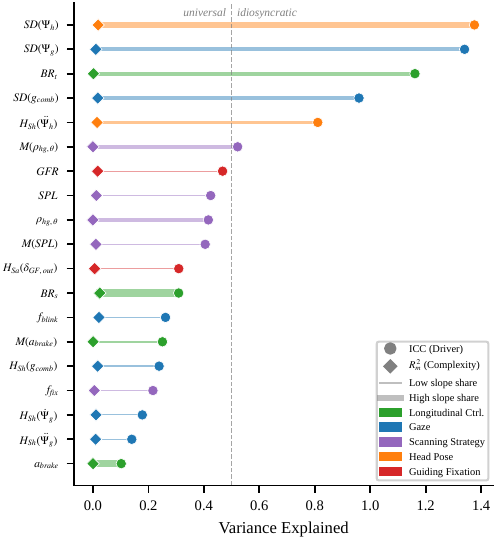}
    \caption{Variance structure of retained features with ICC~$\geq 0.1$. Each row shows ICC (driver identity, $\bullet$) and $R^2_m$ (complexity, $\blacklozenge$) connected by a line whose width encodes slope share. Vertical dashed line at ICC~$= 0.5$ separates universal from idiosyncratic regimes. The remaining 12 features (ICC~$< 0.1$) have near-zero driver and complexity variance.}
    \label{fig:variance_structure}
\end{figure}

The right-hand columns of Table~\ref{tab:screening_variance} report the mixed-effects decomposition for all 31 retained features. Figure~\ref{fig:variance_structure} visualizes the relationship between ICC and $R^2_m$.

Complexity explains 1.5\% of feature variance (median $R^2_m = 0.015$), while driver identity accounts for approximately 23\% (median ICC~$= 0.228$). The remaining $\sim$75\% is within-driver residual variance. The ICC-to-$R^2_m$ ratio of approximately 15:1 confirms that between-driver variance dominates the complexity signal by more than an order of magnitude. Features divide into 25 universal (ICC~$< 0.5$) and six idiosyncratic (ICC~$\geq 0.5$) regimes, the latter concentrated in dispersion-based and head pose metrics (Table~\ref{tab:screening_variance}). At the domain level, braking features exhibit the lowest median ICC (0.064) and highest $R^2_m$ (0.019), while head pose features have the highest median ICC (1.09). The median slope share is 0.196, indicating that approximately 20\% of between-driver variance reflects differences in complexity \textit{sensitivity} (slope) rather than baseline level (intercept). Braking entropy features exhibit the highest slope shares (0.38--0.60), while guiding fixation rate has the lowest (0.033).

\subsection{Personalization}
\label{sec:Results_Personalization}
Table~\ref{tab:personalization} summarizes the effect of all eight normalization strategies on mean $\eta^2$. The per-feature $\Delta\eta^2$ pattern is shown in supplementary Figure~\ref{fig:personalization_heatmap}. All eight strategies degrade discrimination relative to the raw baseline (Table~\ref{tab:personalization}). Among causal strategies, calibration-phase mean subtraction (M2) is least harmful ($-18.3\%$), while sliding-window z-score (L4) is most destructive ($-56.1\%$). Mean-subtraction consistently outperforms z-score, and degradation increases monotonically with shorter temporal scopes.

% ============================================================
% TABLE 3: Feature-Level Personalization Strategies
% ============================================================
\begin{table}[!b]
\centering
% \caption{Feature-level normalization strategies and their effect on mean $\eta^2$ across 31 retained features. $\Delta$: change vs.\ L0 baseline. $n_{drop}$: number of features dropping below $\eta^2 = 0.01$.}
\caption{Effect of feature-level normalization strategies on mean $\eta^2$. $\Delta$: change vs.\ L0 baseline. $n_{drop}$: features dropping below $\eta^2 = 0.01$.}
\label{tab:personalization}
% \scriptsize
\setlength{\tabcolsep}{3pt}
\begin{tabular}{ll cc c}
\textbf{Strategy} & \textbf{Description} & \textbf{Mean} \textbf{$\eta^2$} & \textbf{$\Delta$ (\%)} & \textbf{$n_{drop}$} \\
\hline
  L0 & Raw baseline & 0.019 & --- & --- \\
  L1 & Full-drive z-score & 0.018 & -0.5 & 4 \\
  L2 & Calibration z-score (3~min) & 0.012 & -33.1 & 12 \\
  L3 & Expanding-window z-score & 0.014 & -24.6 & 17 \\
  L4 & Sliding-window z-score (5~min) & 0.008 & -56.1 & 23 \\
  \hline
  M1 & Full-drive mean-sub & 0.018 & -1.1 & 5 \\
  M2 & Calibration mean-sub (3~min) & 0.015 & -18.3 & 11 \\
  M3 & Expanding-window mean-sub & 0.015 & -19.2 & 18 \\
  M4 & Sliding-window mean-sub (5~min) & 0.010 & -47.8 & 23 \\
\end{tabular}
\end{table}

\subsection{Classification}
\label{sec:Results_Classification}
Table~\ref{tab:classification} reports Macro F1 scores for all six models under two cross-validation strategies (see also supplementary Figure~\ref{fig:classification}).

% TABLE 3
% ============================================================
% TABLE 4: Classification Results
% ============================================================
\begin{table}[!b]
\centering
\caption{Classification results: Macro F1~($\pm$~SD) across folds. Gap~=~BlockGroup~$-$~LOSO in percentage points.}
\label{tab:classification}
% \scriptsize
% \setlength{\tabcolsep}{3pt}
\begin{tabular}{l cc c}
\textbf{Model} & \textbf{LOSO} & \textbf{BlockGroup} & \textbf{Gap (pp)} \\
\hline
  Gradient Boosting & 0.452~$\pm$~0.07 & 0.657~$\pm$~0.02 & +20.5 \\
  SVM & 0.428~$\pm$~0.04 & 0.752~$\pm$~0.01 & +32.4 \\
  GRU+Attention & 0.424~$\pm$~0.05 & 0.649~$\pm$~0.01 & +22.5 \\
  Log.~Regression & 0.422~$\pm$~0.07 & 0.464~$\pm$~0.01 & +4.2 \\
  PCA & 0.400~$\pm$~0.04 & 0.402~$\pm$~0.01 & +0.2 \\
  Majority class & 0.210~$\pm$~0.01 & 0.210~$\pm$~0.00 & 0.0 \\
\end{tabular}
\end{table}

Under \ac{LOSO} (primary evaluation), Gradient Boosting achieves the highest Macro F1 of $0.452 \pm 0.07$, followed by SVM ($0.428 \pm 0.04$), GRU+Attention ($0.424 \pm 0.05$), and Logistic Regression ($0.422 \pm 0.08$). All four models converge within a three percentage-point band ($0.422$--$0.452$); the PCA baseline achieves $0.400 \pm 0.04$. The cross-driver generalization gap is substantial: GB drops from $0.657$ (BlockGroupKFold) to $0.452$ (\ac{LOSO}), a difference of 20.5~pp, while SVM exhibits the largest gap (+32.4~pp).

The confusion matrices (supplementary Figure~\ref{fig:confusion_matrices}) reveal that under \ac{LOSO}, the model struggles primarily with the normal class (37\% recall), frequently confusing it with both simple (35\%) and complex (28\%). Simple and complex classes are better separated (62\% and 49\% recall, respectively). Under BlockGroupKFold, all three classes achieve substantially higher recall (78\%, 56\%, 70\%).

\subsection{Feature Importance via SHAP}
\label{sec:Results_SHAP}
\ac{SHAP} values computed on the GB model under \ac{LOSO} CV reveal the multivariate predictive importance of each feature (Figure~\ref{fig:shap_importance}, full ranking in supplementary Table~S1).

\begin{figure}[!htbp]
    \centering
    \includegraphics[width=\columnwidth]{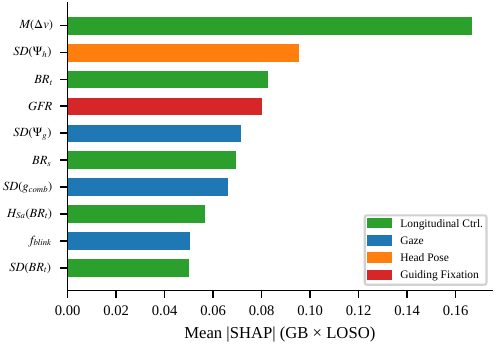}
    \caption{Top ten features by mean $|\text{SHAP}|$ (GB $\times$ LOSO). Color encodes behavioral domain. Full ranking in supplementary Table~\ref{tab:shap}.}
    \label{fig:shap_importance}
\end{figure}

The top-ranked \ac{SHAP} feature is rolling mean speed-limit deviation ($M(\Delta v)$, mean~$|\text{SHAP}| = 0.167$), followed by head yaw dispersion ($SD(\Psi_h)$, $0.095$), braking intensity ($BR_t$, $0.083$), guiding fixation rate ($GFR$, $0.080$), and gaze yaw dispersion ($SD(\Psi_g)$, $0.072$). These top five features span four of five behavioral domains and account for 42\% of total \ac{SHAP} importance. Notably, the \ac{SHAP} ranking diverges from the univariate screening ranking: $M(\Delta v)$ ranks first by \ac{SHAP} but only seventh by $\eta^2$, while $H_{Sa}(BR_t)$ (highest $\eta^2 = 0.046$) ranks eighth by \ac{SHAP}. Three of the top five \ac{SHAP} features are idiosyncratic (ICC~$\geq 0.5$): $SD(\Psi_h)$, $SD(\Psi_g)$, and $BR_t$. Per-feature importance ranks domains as: head pose ($0.065$), longitudinal control ($0.044$), gaze ($0.033$), scanning strategy ($0.031$), and guiding fixation ($0.022$).

\section{Discussion}
\label{sec:Discussion}
The presented multi-stage analysis pipeline screened 175 behavioral features from 20 drivers in real urban traffic, decomposed their variance structure, tested eight personalization strategies, and benchmarked four classification architectures. The following subsections interpret these results along the four research questions, derive deployment implications, and identify limitations.

\subsection{Behavioral Sensitivity to Traffic Complexity}
\label{sec:Disc_RQ1}
Research question RQ1 asked which behavioral features are sensitive to traffic complexity. The complete feature screening retains 31 of 175 features (17 speed-robust, 14 speed-mediated) and identifies four structural patterns that substantially revise and extend the findings of our preceding conference publication~\cite{Koening2026ITSC}.

First, complexity modulates the temporal structure of driver behavior, not steady-state averages. Entropy-based derivations dominate the retained feature set, while gradient and raw-mean derivations are universally null. Our conference publication already identified entropy-based gaze and guiding fixation metrics as novel complexity indicators, but the present systematic screening across 175 features confirms that this pattern holds universally across all five behavioral domains. This extends the finding of Boer~\cite{Boer2001} that behavioral entropy captures control performance variation, and is consistent with Shiferaw et al.~\cite{Shiferaw2019}, who identified entropy-based gaze metrics as more sensitive to task demand than distributional measures. 
Complexity thus manifests as altered behavioral regularity over short time scales rather than shifts in average level, driven by the unpredictable demands of real urban traffic.
% Complexity thus manifests as altered regularity of behavior over short time scales, not as shifts in average level or rate of change. The underlying mechanism is that complex traffic generates unpredictable stimuli, such as merging vehicles, crossing pedestrians, and signal changes, that demand varied behavioral responses, producing higher temporal irregularity in control inputs and gaze patterns.

Second, the behavioral domains differ markedly in sensitivity, with longitudinal control dominating and head pose contributing least (Table~\ref{tab:screening_variance}). Longitudinal control features achieve the highest hit rate and strongest effect sizes, reflecting the most direct behavioral response to complexity. Drivers reduce situational dynamics through speed reduction and anticipatory braking, thereby gaining perceptual time for the increased information-processing demands of complex traffic~\cite{Paxion2014}. This mechanism aligns with Fuller's TCI model~\cite{Fuller2000}. Head pose features, by contrast, are dominated by stable inter-driver differences that mask the complexity signal, compounded by the coarser spatial resolution of head movements relative to eye movements~\cite{Mikula2020,Jha2023}.
% Head pose features, by contrast, are dominated by stable inter-driver differences, i.e., driver identity masks the comparatively small complexity signal. The coarser spatial resolution of head movements relative to eye movements compounds this effect, as small complexity-driven gaze shifts may not propagate to measurable head rotations~\cite{Mikula2020,Jha2023}.

Third, in both gaze and head pose domains a systematic yaw-pitch asymmetry can be observed. All pitch-axis signals are irrelevant, while yaw-axis signals carry complexity information. This contrasts with simulator findings reporting increased vertical dispersion but no change in horizontal dispersion~\cite{Kunst2025,Halin2025}, likely because real urban traffic requires horizontal scanning for intersections, while vertical gaze remains constrained by the fixed dashboard-road geometry.

% Third, in both gaze and head pose domains a systematic yaw-pitch asymmetry can be observed. All pitch-axis signals are irrelevant, while yaw-axis signals carry complexity information. This confirms and extends the pattern observed in our conference publication~\cite{Koening2026ITSC}. Notably, this real-world asymmetry contrasts with the simulator findings of Kunst et al.~\cite{Kunst2025} and Halin et al.~\cite{Halin2025}, who reported increased vertical (pitch) gaze dispersion but no change in horizontal (yaw) dispersion. This discrepancy likely reflects that real urban traffic requires more horizontal scanning for intersections and peripheral monitoring, while vertical gaze remains constrained by the fixed dashboard-road geometry.

Fourth, speed residualization separates two distinct pathways. Speed-mediated features (e.g., speed-limit deviation) lose their effect after regressing out speed, operating entirely through speed adaptation, an ecologically valid response on the causal path from complexity to behavior~\cite{Koening2026ITSC}. Speed-robust features (e.g., braking) retain their full effect, demonstrating that even at matched speeds, drivers brake more frequently and less intensely in complex traffic~\cite{Xie2021_2,Lyu2017}. This mechanistic decomposition is practically relevant for feature selection. % The linear OLS residualization assumes a linear speed-feature relationship; nonlinear associations, if present, would cause the speed-robust category to include partially speed-mediated features, making the classification conservative.

TOST equivalence testing confirms 140 of 143 failing features as null effects. Within guiding fixation, only the continuous guiding fixation rate survives as a speed-robust feature, while all other metrics are null. This indicates that the complexity signal in guiding fixation operates independently of speed adaptation. Mechanistically, complex traffic forces drivers to distribute gaze across peripheral hazard sources, reducing trajectory-aligned guiding fixation~\cite{Lappi2022}.

% The tier ranking shows that the complexity effects identified in the pooled-sample screening are not merely a statistical artifact of inflated sample size. Almost all relevant features also pass the driver-level replication test, meaning that individual drivers show the same directional effect of complexity on these features and they carry valid complexity information.

% Compared with our conference publication~\cite{Koening2026ITSC}, the pipeline confirms 15 of 16 previously identified metrics and adds 16 new features through cross-product construction.

\subsection{Variance Structure: Complexity vs.\ Driver Identity}
\label{sec:Disc_RQ2}
Addressing RQ2, the variance decomposition reveals a 15:1 ratio of driver identity to complexity variance ($\sim$23\% vs.\ $\sim$1.5\%), with residual variance dominating at $\sim$75\%. This makes complexity extraction a fundamentally difficult estimation problem and explains all downstream phenomena: the failure of feature-level personalization (Section~\ref{sec:Disc_RQ3}), the cross-driver generalization gap (Section~\ref{sec:Disc_RQ4}), and the convergence of model architectures at a common performance ceiling. While ICC-based variance partitioning is established in psychometrics, no prior study in the driving domain has quantified this ratio across a multi-domain behavioral feature set, identifying the variance structure as the primary bottleneck for complexity estimation. However, since inter-rater agreement is only moderate, part of the residual variance could reflect label noise and mask complexity variance.

The ICC-based taxonomy separates universal from idiosyncratic features (Table~\ref{tab:screening_variance}). 25 of 31 retained features are universal, meaning that drivers differ in their absolute behavioral levels but respond to complexity in the same direction and with comparable magnitude. 
Speed-limit deviation features, braking entropy features, and guiding fixation rate are precisely the metrics identified in Section~\ref{sec:Disc_RQ1} as most sensitive to complexity, and can be used in cross-driver models without per-driver calibration. 
% Notably, all speed-limit deviation features, braking entropy features, and the guiding fixation rate fall into this group. These are precisely the relevant metrics identified in Section~\ref{sec:Disc_RQ1} as the most sensitive to complexity, and they can be used in cross-driver models without per-driver calibration. 
The six idiosyncratic features concentrate in dispersion-based derivations and the head pose domain matching findings in the literature~\cite{Mikula2020,Arutyunova2025,Jha2023}. They require per-driver calibration for cross-driver use.

The random slope analysis reveals a dissociation between braking and guiding fixation. Braking features have high slope shares, meaning that drivers differ substantially in how they modulate braking under complexity. Guiding fixation rate, by contrast, has the lowest slope share of all retained features, meaning that all drivers reduce it by essentially the same amount under complexity. This uniformity is consistent with the task-imposed nature of guiding fixation and makes guiding fixation rate uniquely suited for driver-agnostic deployment~\cite{Land1994,Lappi2022}. Combining the evidence from research questions RQ1 and RQ2, guiding fixation rate is the only feature that simultaneously satisfies all four deployment criteria, i.e., speed-robustness, Tier~1 replication, universality, and minimal slope share, establishing it as the single most deployment-ready behavioral complexity indicator (Section~\ref{sec:Disc_Implications}).

\subsection{Personalization: Feature-Level Normalization Fails}
\label{sec:Disc_RQ3}

RQ3 tested whether per-driver normalization can recover complexity signal masked by inter-driver differences. The answer is unambiguously negative. All eight normalization strategies degrade complexity discrimination relative to the raw baseline, including the non-causal oracle that represents an upper bound on retrospective normalization. Per-driver z-score normalization is standard practice in driver behavior analysis~\cite{Mariooryad2015,Krejtz2018,Dong2025}. While most prior applications targeted binary state detection tasks with substantially larger effect sizes, for the subtle signal of traffic complexity, the same strategies actively degrade discrimination. Note that z-score normalization rescales feature variance, altering the effect size denominator. However, the identical degradation pattern under mean-subtraction, which preserves the original scale, confirms that this finding is substantive rather than a metric artifact.

The mechanistic explanation follows directly from the variance decomposition: for universal features, no large between-driver offset exists to normalize, while for idiosyncratic features, offset and complexity slope are entangled, so normalization partially removes the signal. Since residual variance dominates at $\sim$75\%, even perfect driver removal recovers only marginal signal.

% Crucially, the direction and magnitude of degradation are not trivially predictable from the variance decomposition alone. That z-score normalization degrades more than mean-subtraction, and that degradation increases monotonically with shorter temporal scopes, are empirical findings that the variance ratios do not predict. The non-causal oracle's near-zero degradation further shows that the failure is not inherent to normalization per se, but to the temporal scarcity of per-driver statistics under causal constraints.

This rules out feature-level personalization as a viable path for complexity-adaptive \ac{DMS}. Model-level approaches are necessary, and the slope share identifies which features benefit most from calibration (e.g. braking entropy) and which can be shared without adjustment (e.g. guiding fixation)~\cite{Lu2023,GLi2024}. %Promising directions include few-shot fine-tuning during a calibration drive, hierarchical Bayesian models with driver-specific random effects, and adversarial domain adaptation treating each driver as a domain~\cite{Lu2023,GLi2024}.

\subsection{Cross-Driver Generalization}
\label{sec:Disc_RQ4}
For RQ4, the classification experiments assessed to what extent complexity can be inferred from driver behavior when the driver is previously unseen. The central result is that complexity estimation from driver behavior is feasible, but performance is fundamentally bounded by the variance structure identified in RQ2. Under \ac{LOSO}, all four model architectures converge within a narrow performance band ($F1 \approx 0.45$), and even the GRU with temporal attention does not break through this ceiling. The convergence confirms that the bottleneck is not model capacity but the 15:1 ratio of driver identity to complexity variance. 
The cross-driver generalization gap precisely matches the ICC-predicted inter-driver barrier.
The confusion matrices reveal that the normal class is the primary source of cross-driver error, as it represents an intermediate regime where behavioral adaptation is partial and less distinct from both adjacent classes.
The SHAP analysis confirms that the GB model distributes importance across all five domains, with guiding fixation rate ranking fourth.
% The \ac{SHAP} analysis confirms that the GB model distributes importance across all five domains, with guiding fixation rate ranking fourth. Three of the top five features are idiosyncratic, partially compensated by the model's non-linearity~\cite{Pena2025}.

\subsection{Implications for Complexity-Adaptive DMS and ADAS}
\label{sec:Disc_Implications}

The findings from research questions RQ1 to RQ4 jointly define three deployment regimes for complexity-adaptive \ac{DMS}, differentiated by the availability of driver-specific calibration data.

The first regime is driver-agnostic deployment. Using only universal features, a population-level GB model achieves $F1 \approx 0.45$ under \ac{LOSO}, which is above the majority-class baseline ($F1 = 0.21$) but insufficient for standalone deployment. This confirms that the complexity signal is extractable from driver behavior alone, while the performance gap quantifies the cost of the unfavorable variance ratio. Here, guiding fixation rate is the single most deployment-ready feature, combining speed-robustness, Tier~1 status, universality, and the lowest slope share and is the only pure gaze metric meeting all four criteria. Combining guiding fixation rate with braking entropy features yields the optimal trade-off between signal strength and universality.
The second regime is personalized deployment, where few-shot model-level calibration could enable safety-relevant applications such as \ac{TOR} timing~\cite{ScharfeScherf2021_2}. 
The third regime is hybrid deployment, combining behavioral features with external context (route complexity priors~\cite{Yang2021,Cheng2022}, V2X traffic density) to push performance beyond the behavioral-only ceiling.

\subsection{Limitations}
\label{sec:Disc_Limitations}

The results are based on a limited set of scenarios due to a sample of 20 individual drivers, unequal gender distribution, urban-only driving on a single route, and driving solely with active, partially automated \ac{ADAS}. The specific variance ratios are dataset-bound, and speed-mediated features may be route-specific due to fixed speed-limit zones. Therefore, the analysis should be validated on a more diverse dataset that includes additional drivers, routes, and environmental conditions. % The method, ICC decomposition combined with multi-strategy cross-validation, generalizes to any driver behavior dataset.

Another limitation concerns the traffic complexity labeling. The expert labels yielded only moderate inter-rater agreement, introducing label noise particularly between normal and complex classes. As a posteriori labeling lacks the real driving dynamics, situational aspects not covered by the video, and the real consequences of actions, it can only serve as an approximation of the real, felt subjective traffic complexity. As noted by Cao et al.~\cite{Cao2025}, a standardized complexity benchmark remains absent, making expert labeling the current best practice. Future studies should complement expert ratings with in-situ driver feedback and validate the labeling approach against objective measures. Furthermore, traffic complexity was labeled using three ordinal classes. A continuous complexity score should be considered to better capture gradual transitions. Because the label definition already included statements about needed adjustments, criterion overlap with the analyzed behavioral metrics cannot be ruled out.

The $\eta^2$-based screening uses the full dataset rather than nested within the cross-validation loop. Since it is a univariate population-level filter with a theory-driven threshold, the risk of overfitting is minimal.

SAE Level~2 automation with fully shared longitudinal control influences braking and speed behavior~\cite{Illgner2024,Koening2026}. The observed braking effects may partly reflect driver-\ac{ADAS} interaction rather than purely voluntary behavioral adaptation. This confound is inherent to the deployment context, as production \ac{DMS} will operate under the same \ac{ADAS} influence, but limits transferability to manual driving scenarios.

Random-slope estimation with only 20 groups limits statistical power for the 2$\times$2 random-effects covariance matrix, and one feature ($a_{brake}$) did not converge. A sensitivity check confirmed that random-intercept-only models yield nearly identical ICC rankings (Spearman $\rho = 0.99$, 2/30 regime changes), indicating robustness to model specification. A gender-balanced replication would allow separating gender effects from idiosyncratic driver differences. Only the first \ac{ADAS}-assisted drive per participant is analyzed, leaving learning effects, habituation, and the effects of traffic complexity on inattentive drivers unexamined.
\section{Conclusion}
\label{sec:Conclusion}
This paper presented an end-to-end analysis of driver behavior as an indicator of traffic complexity in real urban traffic with partially automated \ac{ADAS}, spanning systematic feature screening, variance decomposition, personalization testing, and cross-driver classification.

% The screening of 175 behavioral features across five domains retained 31 with genuine complexity effects and confirmed 140 as null. Entropy-based derivations dominate while gradients and raw means are universally null, longitudinal control is the most sensitive domain, yaw-pitch asymmetry separates relevant from irrelevant signals, and speed residualization distinguishes two mechanistically distinct pathways. 
The screening retained 31 of 175 features with genuine complexity effects, with entropy-based derivations dominating, longitudinal control as the most sensitive domain, a yaw-pitch asymmetry separating relevant from irrelevant signals, and speed residualization distinguishing two mechanistically distinct pathways.
The variance decomposition revealed a 15:1 ratio of driver identity to complexity variance, constituting the fundamental bottleneck. All eight feature-level personalization strategies degraded discrimination, and all four model architectures converged at $F1 \approx 0.45$ under \ac{LOSO}, confirming that this ceiling is imposed by the variance structure.
Across all analysis stages, guiding fixation rate emerged as the single most deployment-ready behavioral complexity indicator. It is the only feature that simultaneously satisfies speed-robustness, driver-level replication, universality across drivers, and minimal inter-driver variation in complexity sensitivity.

These findings carry direct implications for \ac{ADAS} design. The complexity signal is extractable from driver behavior using only production-grade \ac{DMS} sensors, but the unfavorable variance ratio limits standalone deployment. Combining behavioral features with external context, such as route complexity priors or traffic density, offers the most promising path toward complexity-adaptive \ac{ADAS}. The slope share analysis provides a principled basis for selecting which features benefit from per-driver calibration and which can be shared across drivers without adjustment.

Future work should validate these findings on larger, more diverse datasets and evaluate binary complexity classification. 
% Future work should validate these findings on larger, more diverse datasets covering additional routes, driver demographics, and driving conditions and binary complexity classification should be evaluated. 
Model-level personalization through few-shot calibration and hierarchical Bayesian approaches should be investigated as alternatives to the feature-level strategies evaluated here.

\section*{Acknowledgments}
In this paper generative artificial intelligence (GitHub Copilot (Claude Opus 4.6)) was used for code generation, result summary, and text improvements. The authors reviewed and edited the content as needed and take full responsibility for the content of the published article.

%{\appendix[Example Appendix]}

%{\appendices
%\section*{Proof of the First Zonklar Equation}
%Appendix one text goes here.
% You can choose not to have a title for an appendix if you want by leaving the argument blank
%\section*{Proof of the Second Zonklar Equation}
%Appendix two text goes here.}

% References
% \todo[inline]{ITSC Paper PrePrint DOI}
\bibliographystyle{IEEEtran}
\bibliography{root}

\section{Biography Section}
% If you have an EPS/PDF photo (graphicx package needed), extra braces are needed around the contents of the optional argument to biography to prevent the LaTeX parser from getting confused when it sees the complicated $\backslash${\tt{includegraphics}} command within an optional argument. (You can create your own custom macro containing the $\backslash${\tt{includegraphics}} command to make things simpler here.)
 
% \vspace{11pt}

% \bf{If you include a photo:}\vspace{-33pt}
\vspace{-33pt}
\begin{IEEEbiography}[{\includegraphics[width=1in,height=1.25in,clip,keepaspectratio]{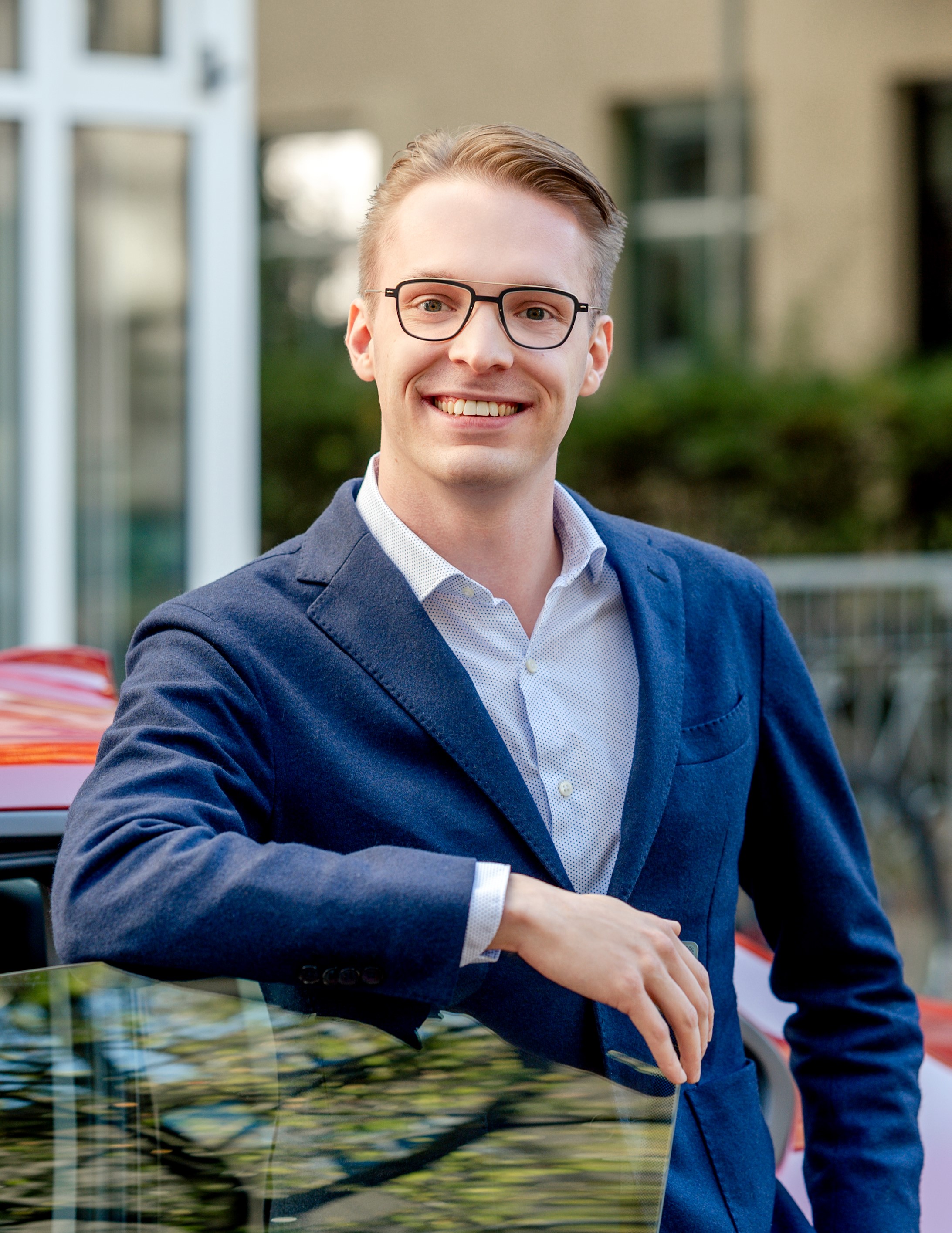}}]{Lukas Köning}
received the B.Sc. degree and the M.Sc. degree in Automotive Engineering from the Technical University of Munich (TUM), in 2022 and 2024, respectively. He is currently pursuing the Ph.D. degree at the Chair of Traffic Engineering and Control (TUM) in cooperation with BMW AG. He has held research and engineering positions at TUM, BMW AG and BMW of North America LLC. His research interests include driver behavior modeling, traffic complexity estimation, and cooperative advanced driver assistance systems.
\end{IEEEbiography}
\vspace{-33pt}
\begin{IEEEbiography}[{\includegraphics[width=1in,height=1.25in,clip,keepaspectratio]{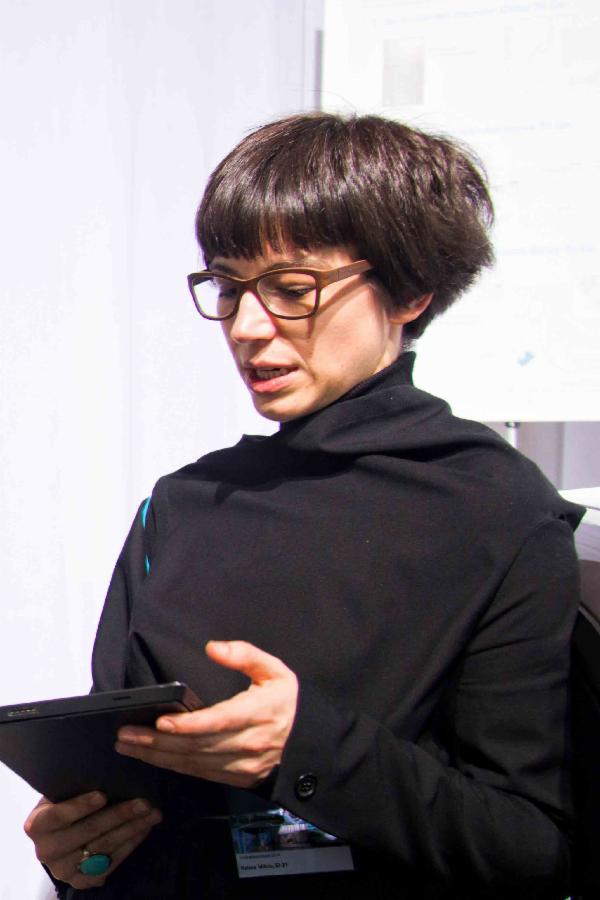}}]{Nataša Miličić}
received the Dr. Dipl.-Ing. degree in Electrical Engineering and Information Technology from the Technical University of Munich (TUM). She is currently Cluster Product Owner for Assisted Driving Level 2 at BMW AG, Munich, Germany, where she is responsible for the customer-centric conception of advanced driver assistance systems. She has held leading positions in product management and human–machine interface development for automated driving. She has contributed to research on automated driving and human–machine interaction. Her research interests include driver behavior, user-centered design, and human–machine interaction for cooperative and automated driving systems.
\end{IEEEbiography}
\vspace{-33pt}
\begin{IEEEbiography}[{\includegraphics[width=1in,height=1.25in,clip,keepaspectratio]{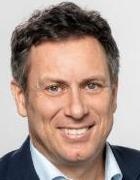}}]{Klaus Bogenberger}
received the Diploma degree in civil engineering and the Ph.D. degree in traffic engineering from the Technical University of Munich, Germany, in 1996 and 2001, respectively. He was a Research Engineer with BMW Group from 2001 to 2008. At first, he was responsible for the traffic flow theory and models with the Department of Science and Transportation and later on, he was with the Department of Corporate Quality. From 2008 to 2011, he was the Managing Director and a Partner of TRANSVER GmbH (consultant office for transport planning and traffic engineering), Munich and Hannover. From 2012 to 2019, he was with Bundeswehr University Munich, as a Professor of traffic engineering. Since 2020, he has been the Chair of Traffic Engineering and Control and the head of the Mobility System Engineering Department, Technical University of Munich. His main research interests include on demand systems, tradeable travel credits, and autonomous vehicles.
\end{IEEEbiography}

\vfill

\newpage
\setcounter{figure}{0}
\setcounter{table}{0}
\renewcommand{\thefigure}{S\arabic{figure}}
\renewcommand{\thetable}{S.\Roman{table}}

\textbf{Supplementary Material}
% ============================================================
% FIGURE S1: Personalization Heatmap
% ============================================================
\begin{figure}[!htbp]
    \centering
    \includegraphics[width=\columnwidth]{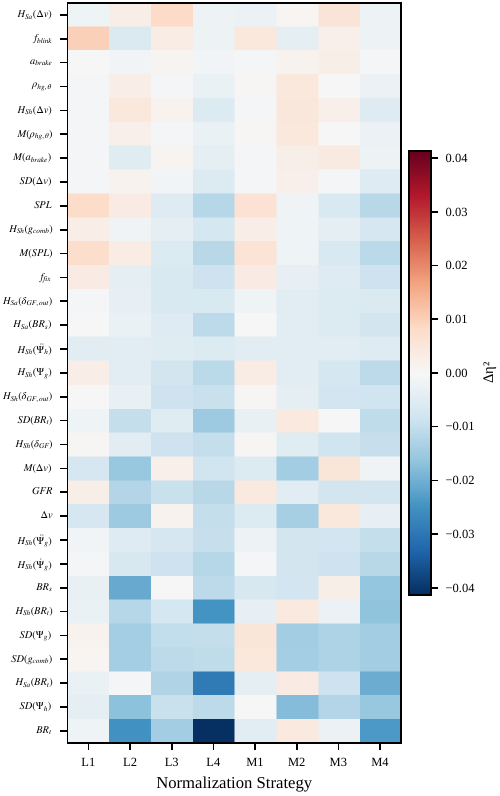}
    \caption{Change in $\eta^2$ ($\Delta\eta^2$) per feature and normalization strategy relative to raw baseline (L0). Blue indicates degradation, red indicates improvement.}
    \label{fig:personalization_heatmap}
\end{figure}

% ============================================================
% FIGURE S2: Classification Performance
% ============================================================
\begin{figure}[!htbp]
    \centering
    \includegraphics[width=\columnwidth]{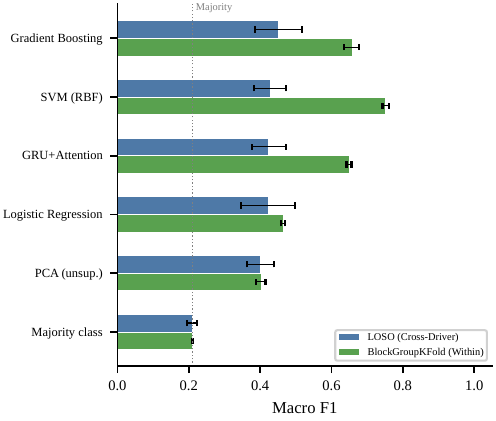}
    \caption{Classification performance (Macro F1) for six models under two cross-validation strategies. Error bars indicate $\pm$SD across folds. Vertical dotted line marks the majority-class baseline. LOSO: leave-one-driver-out, Block: BlockGroupKFold, Strat: StratifiedKFold.}
    \label{fig:classification}
\end{figure}

% ============================================================
% FIGURE S3: Confusion Matrices
% ============================================================
\begin{figure*}[!htbp]
    \centering
    \includegraphics[width=\textwidth]{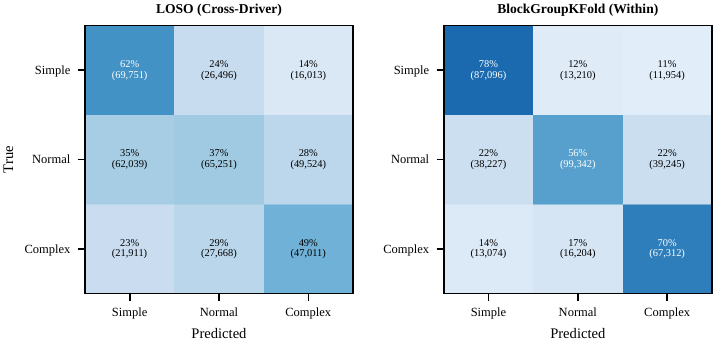}
    \caption{Normalized confusion matrices for the GB model under two cross-validation strategies.}
    \label{fig:confusion_matrices}
\end{figure*}

% ============================================================
% TABLE S4: Detailed Screening Statistics (32 features)
% ============================================================
\begin{table*}[!htbp]
\centering
\caption{Detailed screening statistics for 32 features passing Kruskal-Wallis screening ($\eta^2 \geq 0.01$, $p_{\text{adj}} < .05$). $H$: Kruskal-Wallis test statistic. $p_{\text{adj}}$: Benjamini-Hochberg corrected $p$-value. Dunn $p$: pairwise Dunn post-hoc comparisons (S--M, S--C, M--C). $p_{\text{res}}$: $p$-value of residualized Kruskal-Wallis test. $\eta^2_d$: driver-level effect size ($N=60$). $p_d$: driver-level $p$-value. Regime: universal (U, ICC~$<$~0.5) or idiosyncratic (I, ICC~$\geq$~0.5).}
\label{tab:screening_stats}
\begin{tabular}{c r c ccc c cc c}
\textbf{Feature} & \textbf{$H$} & \textbf{$p_{\text{adj}}$} & \textbf{Dunn $p_{\text{S-M}}$} & \textbf{Dunn $p_{\text{S-C}}$} & \textbf{Dunn $p_{\text{M-C}}$} & \textbf{$p_{\text{res}}$} & \textbf{$\eta^2_d$} & \textbf{$p_d$} & \textbf{Reg.} \\
\hline
  \textbf{Longitudinal Control} \\
  $BR_t$\,[\si{\percent}] & 31717 & $<$.001 & $<$.001 & $<$.001 & $<$.001 & $<$.001 & 0.025 & $<$.001 & I \\
  $H_{Sa}(BR_t)$\,[$-$] & 28415 & $<$.001 & $<$.001 & $<$.001 & $<$.001 & $<$.001 & 0.309 & $<$.001 & U \\
  $H_{Sh}(BR_t)$\,[$-$] & 19126 & $<$.001 & $<$.001 & $<$.001 & $<$.001 & $<$.001 & 0.062 & .138 & U \\
  $M(a_{brake})$\,[\si{\meter/\second\squared}] & 1010 & $<$.001 & $<$.001 & $<$.001 & $<$.001 & $<$.001 & 0.034 & .377 & U \\
  $a_{brake}$\,[\si{\meter/\second\squared}] & 775 & $<$.001 & $<$.001 & $<$.001 & $<$.001 & $<$.001 & 0.034 & .377 & U \\
  $M(\Delta v)$\,[\si{\percent}] & 15917 & $<$.001 & $<$.001 & $<$.001 & $<$.001 & $<$.001 & 0.438 & $<$.001 & U \\
  $\Delta v$\,[\si{\percent}] & 13419 & $<$.001 & $<$.001 & $<$.001 & $<$.001 & $<$.001 & 0.406 & $<$.001 & U \\
  $BR_s$\,[\si{\percent}] & 17177 & $<$.001 & $<$.001 & $<$.001 & $<$.001 & $<$.001 & 0.039 & $<$.001 & U \\
  $H_{Sa}(\Delta v)$\,[$-$] & 6556 & $<$.001 & $<$.001 & $<$.001 & $<$.001 & $<$.001 & 0.221 & $<$.001 & U \\
  $SD(BR_t)$\,[\si{\percent}] & 22812 & $<$.001 & $<$.001 & $<$.001 & $<$.001 & $<$.001 & 0.144 & $<$.001 & U \\
  $H_{Sa}(BR_s)$\,[$-$] & 13000 & $<$.001 & $<$.001 & $<$.001 & $<$.001 & $<$.001 & 0.178 & .004 & U \\
  $H_{Sh}(\Delta v)$\,[$-$] & 6560 & $<$.001 & $<$.001 & $<$.001 & $<$.001 & $<$.001 & 0.375 & $<$.001 & U \\
  $SD(\Delta v)$\,[\si{\percent}] & 9792 & $<$.001 & $<$.001 & $<$.001 & $<$.001 & $<$.001 & 0.475 & $<$.001 & U \\
  \hline
  \textbf{Gaze Behavior} \\
  $f_{blink}$\,[\si{1/\minute}] & 6067 & $<$.001 & $<$.001 & $<$.001 & $<$.001 & $<$.001 & 0.043 & .271 & U \\
  $H_{Sh}(\Psi_g)$\,[$-$] & 6304 & $<$.001 & $<$.001 & $<$.001 & .048 & $<$.001 & 0.163 & .012 & U \\
  $H_{Sh}(\dot{\Psi}_g)$\,[$-$] & 6233 & $<$.001 & $<$.001 & $<$.001 & .176 & $<$.001 & 0.113 & .055 & U \\
  $SD(g_{comb})$\,[\si{\radian}] & 5490 & $<$.001 & $<$.001 & $<$.001 & $<$.001 & $<$.001 & 0.013 & .553 & I \\
  $SD(\Psi_g)$\,[\si{\radian}] & 5759 & $<$.001 & $<$.001 & $<$.001 & $<$.001 & $<$.001 & 0.010 & .718 & I \\
  $H_{Sh}(\ddot{\Psi}_g)$\,[$-$] & 5495 & $<$.001 & $<$.001 & $<$.001 & $<$.001 & $<$.001 & 0.102 & .036 & U \\
  $H_{Sh}(g_{comb})$\,[$-$] & 4478 & $<$.001 & $<$.001 & $<$.001 & .255 & $<$.001 & 0.079 & .064 & U \\
  \hline
  \textbf{Head Pose} \\
  $SD(\Psi_h)$\,[\si{\radian}] & 8150 & $<$.001 & $<$.001 & $<$.001 & $<$.001 & $<$.001 & 0.027 & .448 & I \\
  $H_{Sh}(\ddot{\Psi}_h)$\,[$-$] & 3378 & $<$.001 & $<$.001 & $<$.001 & $<$.001 & $<$.001 & 0.016 & .744 & I \\
  \hline
  \textbf{Guiding Fixation} \\
  $GFR$\,[\si{\percent}] & 14683 & $<$.001 & $<$.001 & $<$.001 & $<$.001 & $<$.001 & 0.070 & $<$.001 & U \\
  $H_{Sh}(\delta_{GF})$\,[$-$] & 7158 & $<$.001 & $<$.001 & $<$.001 & $<$.001 & $<$.001 & 0.218 & .002 & U \\
  $H_{Sh}(\delta_{GF,out})$\,[$-$] & 7047 & $<$.001 & $<$.001 & $<$.001 & $<$.001 & $<$.001 & 0.190 & $<$.001 & U \\
  $H_{Sa}(\delta_{GF,out})$\,[$-$] & 6855 & $<$.001 & $<$.001 & $<$.001 & $<$.001 & $<$.001 & 0.047 & .002 & U \\
  $H_{Sh}(t_{GF})$\,[$-$] & 5155 & $<$.001 & $<$.001 & $<$.001 & $<$.001 & $<$.001 & 0.000 & .996 & --- \\
  \hline
  \textbf{Scanning Strategy} \\
  $SPL$\,[\si{\degree}] & 5901 & $<$.001 & $<$.001 & $<$.001 & $<$.001 & $<$.001 & 0.018 & .466 & U \\
  $M(SPL)$\,[\si{\degree}] & 5135 & $<$.001 & $<$.001 & $<$.001 & $<$.001 & $<$.001 & 0.013 & .510 & U \\
  $M(\rho_{hg,\theta})$\,[$-$] & 4282 & $<$.001 & $<$.001 & $<$.001 & $<$.001 & $<$.001 & 0.021 & .382 & I \\
  $\rho_{hg,\theta}$\,[$-$] & 3942 & $<$.001 & $<$.001 & $<$.001 & $<$.001 & $<$.001 & 0.022 & .413 & U \\
  $f_{fix}$\,[\si{\hertz}] & 5685 & $<$.001 & $<$.001 & $<$.001 & $<$.001 & $<$.001 & 0.070 & .141 & U \\
\end{tabular}
\end{table*}

% ============================================================
% TABLE S2: TOST Equivalence Summary by Domain
% ============================================================
\begin{table}[!t]
\centering
\caption{TOST equivalence-confirmed null features by domain. 140 of 175 features receive formal equivalence confirmation. $p_{\text{TOST}}$: maximum across all six one-sided tests. $n_{BF>10}$: features with Bayes Factor $BF_{01} > 10$ (strong evidence for null).}
\label{tab:tost_null}
\begin{tabular}{l r c c r}
\textbf{Domain} & \textbf{$n_{\text{null}}$} & \textbf{$\eta^2$ range} & \textbf{$p_{\text{TOST}}$} & \textbf{$n_{BF>10}$} \\
\hline
  Gaze Behavior & 35 & 0.000--0.008 & all $<$.05 & 3 \\
  Head Pose & 40 & 0.000--0.007 & all $<$.05 & 4 \\
  Guiding Fixation & 28 & 0.000--0.009 & all $<$.05 & 7 \\
  Scanning Strategy & 29 & 0.000--0.009 & all $<$.05 & 6 \\
  Longitudinal Control & 8 & 0.000--0.006 & all $<$.05 & 1 \\
  \hline
  \textbf{Total} & \textbf{140} & 0.000--0.009 & 138: $<$.001 & \textbf{21} \\
\end{tabular}
\end{table}

% ============================================================
% TABLE S1: SHAP Feature Importance (all 31 features)
% ============================================================
\begin{table}[!t]
\centering
\caption{SHAP feature importance for all 31 retained features (Gradient Boosting~$\times$~LOSO cross-validation): mean $|\text{SHAP}|$ averaged across 20 folds.}
\label{tab:shap}
\begin{tabular}{rl c c c c}
\textbf{Rank} & \textbf{Feature} & \textbf{Domain} & \textbf{Mean $|\text{SHAP}|$} & \textbf{$\eta^2$} & \textbf{ICC} \\
\hline
  1 & $M(\Delta v)$\,[\si{\percent}] & Long.Ctrl & 0.167 & 0.028 & 0.044 \\
  2 & $SD(\Psi_h)$\,[\si{\radian}] & Head & 0.095 & 0.020 & 1.376 \\
  3 & $BR_t$\,[\si{\percent}] & Long.Ctrl & 0.083 & 0.057 & 1.161 \\
  4 & $GFR$\,[\si{\percent}] & Guid.Fix & 0.080 & 0.029 & 0.468 \\
  5 & $SD(\Psi_g)$\,[\si{\radian}] & Gaze & 0.072 & 0.015 & 1.340 \\
  6 & $BR_s$\,[\si{\percent}] & Long.Ctrl & 0.070 & 0.025 & 0.309 \\
  7 & $SD(g_{comb})$\,[\si{\radian}] & Gaze & 0.066 & 0.015 & 0.960 \\
  8 & $H_{Sa}(BR_t)$\,[$-$] & Long.Ctrl & 0.057 & 0.046 & 0.011 \\
  9 & $f_{blink}$\,[\si{\per\minute}] & Gaze & 0.051 & 0.019 & 0.262 \\
  10 & $SD(BR_t)$\,[\si{\percent}] & Long.Ctrl & 0.050 & 0.020 & 0.018 \\
  11 & $SD(\Delta v)$\,[\si{\percent}] & Long.Ctrl & 0.050 & 0.010 & 0.005 \\
  12 & $M(SPL)$\,[\si{\degree}] & Scan.Strat & 0.048 & 0.014 & 0.405 \\
  13 & $f_{fix}$\,[\si{\hertz}] & Scan.Strat & 0.041 & 0.010 & 0.216 \\
  14 & $H_{Sa}(BR_s)$\,[$-$] & Long.Ctrl & 0.037 & 0.016 & 0.064 \\
  15 & $H_{Sh}(\ddot{\Psi}_h)$\,[$-$] & Head & 0.035 & 0.010 & 0.811 \\
  16 & $M(\rho_{hg,\theta})$\,[$-$] & Scan.Strat & 0.034 & 0.012 & 0.522 \\
  17 & $H_{Sh}(\Delta v)$\,[$-$] & Long.Ctrl & 0.033 & 0.015 & 0.013 \\
  18 & $H_{Sh}(\dot{\Psi}_g)$\,[$-$] & Gaze & 0.022 & 0.016 & 0.178 \\
  19 & $SPL$\,[\si{\degree}] & Scan.Strat & 0.019 & 0.016 & 0.425 \\
  20 & $\Delta v$\,[\si{\percent}] & Long.Ctrl & 0.017 & 0.026 & 0.043 \\
  21 & $\rho_{hg,\theta}$\,[$-$] & Scan.Strat & 0.014 & 0.011 & 0.416 \\
  22 & $H_{Sh}(\Psi_g)$\,[$-$] & Gaze & 0.010 & 0.016 & 0.099 \\
  23 & $H_{Sa}(\Delta v)$\,[$-$] & Long.Ctrl & 0.005 & 0.021 & 0.017 \\
  24 & $H_{Sh}(\ddot{\Psi}_g)$\,[$-$] & Gaze & 0.004 & 0.014 & 0.140 \\
  25 & $H_{Sh}(\delta_{GF,out})$\,[$-$] & Guid.Fix & 0.004 & 0.017 & 0.094 \\
  26 & $M(a_{brake})$\,[\si{\meter\per\second\squared}] & Long.Ctrl & 0.004 & 0.030 & 0.251 \\
  27 & $H_{Sh}(g_{comb})$\,[$-$] & Gaze & 0.004 & 0.011 & 0.239 \\
  28 & $a_{brake}$\,[\si{\meter\per\second\squared}] & Long.Ctrl & 0.003 & 0.028 & 0.102 \\
  29 & $H_{Sh}(\delta_{GF})$\,[$-$] & Guid.Fix & 0.003 & 0.018 & 0.076 \\
  30 & $H_{Sh}(BR_t)$\,[$-$] & Long.Ctrl & 0.002 & 0.033 & 0.020 \\
  31 & $H_{Sa}(\delta_{GF,out})$\,[$-$] & Guid.Fix & 0.002 & 0.011 & 0.310 \\
\end{tabular}
\end{table}

\end{document}